\documentclass[a4paper,10pt]{article}
\usepackage{graphicx}
\usepackage{amsmath}
\usepackage{amssymb}
\usepackage[dvipsnames]{xcolor}
\usepackage{booktabs,cite,jheppub,multirow}
\hypersetup{linkcolor=Maroon,citecolor=Maroon,filecolor=Maroon,urlcolor=Maroon}
\usepackage{physics}
\usepackage{dsfont}
\usepackage{tikz}
\usepackage{xspace}

\usepackage{cleveref}
\crefname{section}{Section}{Sections}
\crefname{table}{Table}{Tables}
\crefname{figure}{Figure}{Figures}
\crefname{equation}{Eq.}{Eqs.}
\crefname{appendix}{Appendix}{Appendix}

\newcommand{\SO}{\mathrm{SO}}
\newcommand{\SU}{\mathrm{SU}}
\newcommand{\U}{\mathrm{U}}
\newcommand{\Sp}{\mathrm{Sp}}

\def\madanalysis{{\texttt{MadAnalysis5}}\xspace}
\def\hackanalysis{{\texttt{HackAnalysis}}\xspace}
\def\pyhf{{\texttt{pyhf}}\xspace}
\def\spey{{\texttt{spey}}\xspace}

\title{Composite top partners in exotic colour representations}

\author[a]{\! Giacomo Cacciapaglia,}
\affiliation[a]{Laboratoire de Physique Théorique et Hautes Énergies (LPTHE), UMR 7589, Sorbonne Université et CNRS, 4 place Jussieu, 75252 Paris Cedex 05, France}
\emailAdd{cacciapa@lpthe.jussieu.fr}
\author[b]{Rosy Caliri,}
\affiliation[b]{Institut f\"{u}r Theoretische Physik und Astrophysik, Uni W\"{u}rzburg, Emil-Hilb-Weg 22, D-97074 W\"{u}rzburg, Germany}
\emailAdd{rosy.caliri@uni-wuerzburg.de}
\author[c,d]{\! Aldo Deandrea,}
\affiliation[c]{Universit\'e Lyon 1, CNRS, IP2I, UMR 5822, Villeurbanne, France}
\affiliation[d]{Department of Physics, University of Johannesburg, 
PO Box 524, Auckland Park 2006, South Africa}
\emailAdd{deandrea@ip2i.in2p3.fr}
\author[a]{\! Benjamin Fuks,}
\emailAdd{fuks@lpthe.jussieu.fr}
\author[a]{\! Mark Goodsell,}
\emailAdd{goodsell@lpthe.jussieu.fr}
\author[b]{\! Jan~Hadlik,}
\emailAdd{jan.hadlik@uni-wuerzburg.de}
\author[b]{\! Manuel Kunkel}
\emailAdd{manuel.kunkel@uni-wuerzburg.de}
\author[b]{\! and \! Werner Porod}
\emailAdd{werner.porod@uni-wuerzburg.de}

\abstract{
  Composite Higgs models with partial compositeness generically predict coloured fermi\-onic resonances associated with the strong dynamics responsible for electroweak symmetry breaking. While most phenomenological studies have focused on colour-triplet and colour-octet top partners, several UV-complete hypercolour constructions also contain fermionic colour sextets. We present a systematic study of these states in the minimal model classes where they arise, constructing the relevant low-energy interactions and deriving their characteristic decay patterns. The sextets predominantly decay through coloured pseudo-Nambu-Goldstone bosons, leading to top-rich final states, while additional channels with $b$-jets and missing transverse energy can be important. We reinterpret existing ATLAS and CMS searches for high-multiplicity final states to derive the dedicated constraints on these resonances. For the benchmark spectra considered, current LHC data exclude individual sextet components up to masses in the $2-2.5$~TeV regime, with stronger bounds when the full sextet multiplet is included, while conservative extrapolations to the HL-LHC indicate a reach close to 3~TeV. Our results therefore show that colour-sextet fermions provide a powerful and largely unexplored probe of composite Higgs models with partial compositeness.
}

\begin{document}

\maketitle

\section{Introduction}

Composite Higgs models offer one of the few known mechanisms for a symmetry-based explanation of the electroweak (EW) hierarchy problem of the Standard Model (SM). The spontaneous symmetry breaking is realised via confining dynamics~\cite{Weinberg:1975gm,Dimopoulos:1979es,Eichten:1979ah}, and the first models, inspired by QCD, did not feature a light Higgs boson. It was soon realised that a Higgs boson could emerge as a pseudo-Nambu-Goldstone boson
(pNGB) from vacuum misalignment in the breaking of the associated global symmetry \cite{Kaplan:1983fs,Kaplan:1983sm}. The masses of the SM fermions, in turn, can be generated through linear 
couplings between the elementary SM fermion fields and composite operators via the mechanism of partial compositeness \cite{Kaplan:1991dc}.
A more recent realisation of a composite (Goldstone) Higgs relies on five-dimensional holographic model building~\cite{Contino:2003ve,Agashe:2004rs,Agashe:2006at}, where the composite Higgs emerges as a bulk gauge degree of freedom in the same fashion as in gauge-Higgs unification~\cite{Hosotani:1983vn,Haba:2004qf,Hosotani:2005nz}.
Furthermore, an interesting class of models arises from confining four-dimensional gauge theories (hypercolour) with condensing
fermionic matter (hyperfermions), where the symmetry breaking pattern is determined by the gauge representation of the hyperfermions~\cite{Peskin:1980gc,Preskill:1980mz,Cacciapaglia:2014uja}. The condensate breaks a global flavour symmetry $G$ of the hyperfermion sector down to an unbroken subgroup $H$, giving rise to pNGBs that live in the coset $G/H$. In the EW sector, the following minimal cosets are obtained: $G/H=\SU(5)/\SO(5)$ \cite{Ferretti:2014qta,Golterman:2015zwa,Ferretti:2016upr,Agugliaro:2018vsu} for real hyperfermions; 
$\SU(4)/\Sp(4)$ \cite{Barnard:2013zea,Ferretti:2013kya,Ferretti:2014qta,Cacciapaglia:2014uja,Galloway:2016fuo,Agugliaro:2016clv} for pseudo-real ones; 
and $\SU(4)\times \SU(4)/\SU(4)$ \cite{Ma:2015gra,Alanne:2017rrs} for complex ones. 
The phenomenological requirement to realise the Higgs field as a bi-doublet $(\mathbf{2},\mathbf{2})$ of the custodial symmetry 
$\SU(2)_L \times \SU(2)_R \subset H$ fixes the minimal dimension of the above cosets. Generating fermionic resonances requires additional hyperfermions that carry QCD charges, and a full classification of possible models was presented in~\cite{Barnard:2013zea,Ferretti:2013kya,Ferretti:2016upr}. 
We stress that such models should be considered as an underlying description of the low energy dynamics rather than ultra-violet completions for composite Higgs models. In fact, generating the appropriate dynamics above the confinement scale and sourcing the couplings responsible for partial compositeness requires further model-building efforts, see, for instance, in~\cite{Cacciapaglia:2018avr,Cacciapaglia:2019dsq,Cacciapaglia:2019vce,Cacciapaglia:2020jvj}.

The spectrum of the resulting low energy theory at the (multi-)TeV scale consists of states invariant under the hypercolour group $G_\mathrm{HC}$ and built from hyperfermions. Bilinear combinations give rise to pNGBs and spin-1 vector resonances, while trilinear combinations give rise to fermionic bound states. For the latter, it is particularly convenient to employ two species of hyperfermions: one carrying EW charges and the other QCD and hypercharge~\cite{Barnard:2013zea,Ferretti:2013kya}. In this way, the sector responsible for the EW symmetry breaking is sequestered from the one carrying QCD charges, avoiding the potential spontaneous breaking of QCD and the presence of leptoquarks~\cite{Gripaios:2014tna,Vecchi:2015fma,Marzocca:2018wcf}.
The fermionic bound states are commonly
called {\it top partners}, as they are responsible for the generation of the top mass. In practical terms, models with two species of hyperfermions extend the global symmetry group $G$ to a product group, with one factor containing the EW part of the SM group and the other containing the colour part, together with hypercharge~\cite{Barnard:2013zea,Ferretti:2013kya,Ferretti:2016upr}. The minimal requirement for the coloured hyperfermions is to contain one triplet and one anti-triplet with appropriate hypercharges, leading to the following minimal colour-sector cosets: $\SU(6)/\SO(6)$,  $\SU(6)/\Sp(6)$ and $\SU(3)\times \SU(3)/\SU(3)$. Besides giving mass to the top quark, top partners also contribute to the Higgs potential and self-couplings, possibly providing the dominant contribution to the Higgs mass~\cite{Panico:2015jxa,Cacciapaglia:2020kgq}. One-loop estimates often point to a strong upper bound on the top-partner masses of about $2$~TeV~\cite{Panico:2012uw}. However, as top partners are resonances of the strongly interacting hypercolour theory, the loop computation is only valid under limited assumptions~\cite{Contino:2011np}. More general analyses, see \textit{e.g.}~\cite{Alanne:2018wtp,Agugliaro:2018vsu}, leave top-partner masses largely unconstrained, so that masses in the multi-TeV range should also be considered.
This expectation is further supported by non-perturbative studies of candidate hypercolour theories, including lattice computations for $\SU(4)$ models~\cite{Ayyar:2017qdf,Ayyar:2018glg,Ayyar:2018ppa,Ayyar:2018zuk,Ayyar:2019exp,Hasenfratz:2023sqa} and for $\Sp(4)$ models~\cite{Bennett:2017kga,Bennett:2019cxd,Bennett:2019jzz,Bennett:2022yfa,Kulkarni:2022bvh,Bennett:2023mhh,Bennett:2024wda,Bennett:2024tex,TELOS:2025ash}. Furthermore, complementary estimates based on gauge/gravity duality~\cite{Erdmenger:2020lvq,Erdmenger:2020flu,Elander:2020nyd,Elander:2021bmt,Erdmenger:2023hkl,Erdmenger:2024dxf} also point to a broadly consistent picture for such models.

This class of composite (Goldstone) Higgs models based on hypercolour interactions offers a particularly rich and predictive spectrum, since the quantum numbers of the resonances follow directly from the underlying hyperfermion representations.\footnote{In contrast, in holographic constructions, resonances are introduced as bulk fields in the extra dimension, and their quantum numbers are specified by the model builder.} The phenomenology of models with two hyperfermion species has therefore been explored in several directions. The extended electroweak pNGB sector has been studied in \cite{Ferretti:2016upr,Agugliaro:2018vsu,Cacciapaglia:2022bax,Flacke:2023eil}, while the axion-like states associated with the additional spontaneously broken $\U(1)$ global symmetries have been investigated in~\cite{Belyaev:2015hgo, Belyaev:2016ftv, Cacciapaglia:2019bqz, Cornell:2020usb, BuarqueFranzosi:2021kky, Ferretti:2025zsq}. The QCD-coloured pNGBs provide another distinctive set of signatures and have been analysed in  \cite{Cacciapaglia:2015eqa,Belyaev:2016ftv,Cacciapaglia:2020vyf,Flacke:2025xwl}. Heavier resonances have also received significant attention. The spin-1 sector is divided into electroweak resonances~\cite{BuarqueFranzosi:2016ooy,Caliri:2024jdk} and QCD-coloured resonances~\cite{Cacciapaglia:2024wdn}. In both cases, states that mix with the SM gauge bosons are subject to mass bounds in the multi-TeV range.
The fermionic bound states can appear in several colour representations. The colour-triplet states have been studied extensively, for instance in~\cite{DeSimone:2012fs,Buchkremer:2013bha} or in scenarios where they undergo non-standard decays into the additional pNGBs \cite{Bizot:2018tds,Cacciapaglia:2019zmj,Xie:2019gya,Benbrik:2019zdp,Wang:2020ips,Corcella:2021mdl,Banerjee:2022izw,Banerjee:2022xmu,Banerjee:2024zvg,Flacke:2026xyz}.
Current associated LHC searches typically exclude colour-triplet top partners with decays to SM particles up to masses of order $1.5-1.7$~TeV, depending on the electroweak representation and branching ratios~\cite{CMS:2022fck, ATLAS:2022hnn, ATLAS:2024gyc, CMS:2024bni, ATLAS:2024xdc}. By contrast, the non-triplet fermionic resonances predicted by the same underlying constructions are much less explored. Colour-octet top partners have recently been explored in~\cite{Cacciapaglia:2021uqh}, leading to bounds on their masses of up to about $2.7$~TeV from existing LHC data. Colour-sextet fermions, however, remain largely unexplored at the LHC despite being robustly predicted in several hypercolour constructions.

In this paper, we investigate the LHC phenomenology of fermionic colour sextets, which, to the best of our knowledge, have not been systematically studied in the context of Composite Higgs models. Fermionic colour sextets have instead appeared only sparsely in the phenomenological literature~\cite{Chivukula:1990di,Celikel:1998dj,Carpenter:2021rkl}. In hypercolour constructions with partial compositeness, these sextet hyperbaryons are particularly well motivated: they arise in the same strongly coupled sector as the usual colour-triplet top partners, can naturally have masses of the same order, and benefit from enhanced QCD pair-production rates due to their non-minimal colour representation. They therefore provide a potentially powerful probe of the coloured fermionic spectrum. In this work, we systematically investigate the sextet signatures characteristic of these models, including cascade decays through coloured pNGBs and final states with very large top- and bottom-quark multiplicities. We show that existing LHC data already imply bounds on sextet masses up to about $2.6$~TeV, while projections to the High Luminosity LHC (HL-LHC) indicate a reach close to $3$~TeV.

The paper is structured as follows. In \cref{sec:models}, we collect the relevant aspects of the models considered and present the interaction Lagrangian governing the sextet dynamics. In \cref{sec:pheno} we discuss the LHC phenomenology of the fermionic top partners transforming in the sextet representation of $\SU(3)_c$. In \cref{sec:LHC}, we then derive mass bounds on these states from existing LHC searches and discuss the projected sensitivity at the HL-LHC. Finally, in \cref{sec:outlook} we draw our conclusions and present an outlook. Technical details are collected in the appendices.

\section{Modern models of partial compositeness}\label{sec:models}
In this section, we present the class of composite Higgs models that forms the basis of our analysis. We begin in \cref{sec:hyperfermions} by introducing the hyperfermion content and the symmetry-breaking structure of the models, which determine both the pNGB spectrum, which is discussed in \cref{sec:pNGBspectrum}, and the baryonic one that we introduce in \cref{sec:baryonsspectrum}. Since this paper focuses on the phenomenology of the colour-sextet states inherent to these models, we dedicate \cref{sec:sextets} to the origins of their couplings in the framework of partial compositeness. Finally, \cref{sec:M12} is reserved to an outlier model with even more exotic sextet phenomenology, which we leave for further studies.

\subsection{Hyperfermion content and symmetry breaking}\label{sec:hyperfermions}
We consider a class of composite Higgs models that simultaneously account for electroweak symmetry breaking~\cite{Kaplan:1983fs} and provide fermionic bound states suitable for the partial compositeness mechanism to generate a large top-quark mass~\cite{Kaplan:1991dc}. These models were originally proposed in~\cite{Ferretti:2013kya, Ferretti:2016upr} and further developed in~\cite{Golterman:2015zwa, DeGrand:2016pgq, Belyaev:2016ftv}. They are formulated as four-dimensional asymptotically-free gauge theories with a new confining interaction, referred to as hypercolour (HC), which becomes strongly coupled at a scale $\Lambda_{\mathrm{HC}}$. The fundamental degrees of freedom are the so-called hyperfermions, and the absence of elementary scalars ensures that the Higgs sector emerges dynamically, thereby ameliorating the hierarchy problem. A distinctive feature of these constructions is the presence of two species of hyperfermions, denoted by $\psi$ and $\chi$, transforming under different irreducible representations of the hypercolour gauge group $G_{\mathrm{HC}}$. This two-representation structure is essential in order to simultaneously realise a pNGB Higgs sector and composite fermionic operators with the quantum numbers required for partial compositeness, while sequestering QCD charges (carried by the $\chi$ hyperfermions) from the Higgs sector (generated by the $\psi$ hyperfermions). A similar feat could be achieved by one-species models based on the $\SU(3)_{\rm HC}$ \cite{Ferretti:2013kya, Vecchi:2015fma, Marzocca:2018wcf} or $G_2$ \cite{Ferretti:2013kya} groups with fermions in the fundamental representation, at the price of predicting composite leptoquarks \cite{Gripaios:2014tna,Marzocca:2018wcf}.

The $\psi$ hyperfermions are uncoloured but carry EW quantum numbers, and their condensation triggers the spontaneous breaking of a global flavour symmetry $G_F^\psi$ to a subgroup $H_F^\psi$, giving rise to a set of pNGBs among which the Higgs doublet is identified. In realistic constructions, the unbroken subgroup contains a custodial symmetry, thereby ensuring compatibility with electroweak precision tests. The $\chi$ hyperfermions, on the other hand, transform as coloured states of the QCD gauge group $\SU(3)_c$ and carry hypercharge. Their presence enlarges the global flavour symmetry to $G_F = G_F^\psi \times G_F^\chi \times \U(1)_\psi \times \U(1)_\chi$, where $G_F^\chi$ is spontaneously broken by the formation of a $\chi$-condensate at the confinement scale. The two $\U(1)$ global symmetries are also spontaneously broken by the condensates, leading to two axion-like particles in the spectrum: while one is expected to be heavy due to the HC-gauge anomaly of one linear combination of the two independent Abelian symmetries, the other may be very light and lead to an interesting phenomenology~\cite{Belyaev:2015hgo,Belyaev:2016ftv, Cacciapaglia:2017iws, Cacciapaglia:2019bqz, Cornell:2020usb, BuarqueFranzosi:2021kky}. The $\chi$ sector gives rise to coloured pNGBs~\cite{Cacciapaglia:2015eqa,Cacciapaglia:2020vyf}. The separation of electroweak and colour quantum numbers between the $\psi$ and $\chi$ species ensures that EW symmetry breaking is primarily driven by the $\psi$ sector while avoiding vacuum misalignment that would lead to the breaking of the QCD symmetry. At the same time, mixed hyperbaryon states involving both $\psi$ and $\chi$ fields provide the necessary spectrum of coloured fermionic resonances, including top partners that, depending on the model, are found to transform as singlets, triplets, sextets or octets under $\SU(3)_c$~\cite{Cacciapaglia:2021uqh}.

Many choices of the hypercolour gauge symmetry and hyperfermion representations may lead to the above features. However, theoretical requirements strongly constrain the set of possibilities. Firstly, the gauge interactions must confine and generate a mass gap without breaking the hypercolour symmetry: as the models are always vector-like, the latter is ensured by the Vafa-Witten theorem~\cite{Vafa:1983tf}. The generation of the mass gap requires the theories to be outside of the conformal window, hence limiting the hyperfermion multiplicity~\cite{Ferretti:2016upr, Belyaev:2016ftv, Kim:2020yvr}.\footnote{A walking regime is generically required above $\Lambda_{\rm HC}$ in order to generate a hierarchy with the flavour symmetry breaking scale via large anomalous dimensions for the baryons~\cite{BuarqueFranzosi:2019eee}. This can be achieved by adding more hyperfermions with mass around $\Lambda_{\rm HC}$.} The symmetry breaking patterns are then fixed by the nature of the hyperfermion representations being complex~\cite{Vafa:1983tf} or real/pseudo-real~\cite{Kosower:1984aw}. Applying these theoretical consistency conditions together with phenomenological requirements selects a finite subset of twelve viable models \cite{Ferretti:2016upr,Belyaev:2016ftv}, which differ by the choice of the hypercolour gauge group and by the representations and multiplicities of the $\psi$ and $\chi$ fermions. Depending on the representations, fermionic bound states (hyperbaryons) can schematically be constructed either as $\psi\chi\psi$ or as $\chi\psi\chi$ composites. While the former configuration typically yields only colour-triplet fermionic resonances, the latter produces a richer spectrum.  In this work, we focus on this subset of models that we organise into the four classes C1-C4 introduced in \cref{tab:modelclasses}, with each class corresponding to a given spectrum of pNGBs.
As models in classes C1-C3 share many common features, we will discuss their spectrum and signatures at length. The model M12, \textit{i.e.}\ the sole member of class C4, exhibits qualitatively different features, hence it will be treated separately in \cref{sec:M12}. 

\begin{table}\renewcommand{\arraystretch}{1.5} \setlength{\tabcolsep}{6pt}
    \centering\resizebox{\columnwidth}{!}{%
    \begin{tabular}{ccccc|cc}
        \toprule
        Class & Model & $G_{\mathrm{HC}}$ & $\psi$ repr. & $\chi$ repr. & $\psi\psi$ & $\chi\chi$ \\ \midrule
        C1 & M1, M2 & $\mathrm{SO}(7), \mathrm{SO}(9)$ &$5\times \mathbf{F}$ (real) & $6\times \mathbf{Spin}$ (real) 
           & $S^{++}, S^+, S^0$ & $\pi_8, \pi_6$ \\
        C2 & M5     & $\mathrm{Sp}(4)$ &$5\times \mathbf{A}_2$ (real) & $6\times \mathbf{F}$ (pseudo-real)
           & $S^{++}, S^+, S^0$ & $\pi_8, \pi_3$ \\
        \multirow{2}{*}{C3} & M6 & $\mathrm{SU}(4)$ & $5\times \mathbf{A}_2$ (real) & $3\times (\mathbf{F}, \overline{\mathbf{F}})$ (complex)
             & \multirow{2}{*}{$S^{++}, S^+, S^0$} & \multirow{2}{*}{$\pi_8$} \\[-.1cm] 
                            & M7 & $\mathrm{SO}(10)$ & $5\times \mathbf{F}$ (real) & $3\times (\mathbf{Spin}, \overline{\mathbf{Spin}})$ (complex) \\ 
        C4 & M12    & $\mathrm{SU}(5)$ & $4\times (\mathbf{F}, \overline{\mathbf{F}})$ (complex) 
        & $3\times (\mathbf{A}_2, \overline{\mathbf{A}_2})$ (complex) & $S^+, S^0$ & $\pi_8$ \\
        \bottomrule
    \end{tabular}}
    \caption{Field content of the subset of composite Higgs models considered in this work, with the hypercolour gauge group $G_{\mathrm{HC}}$, the representations chosen for the $\psi$ and $\chi$ hyperfermions, together with their multiplicities, and the resulting pNGB spectrum. The model naming scheme follows the classification introduced in~\cite{Belyaev:2016ftv}, and the representations are denoted as  fundamental ($\mathbf{F}$), two‑index antisymmetric ($\mathbf{A}_2$) and spinorial ($\mathbf{Spin}$) (for $\mathrm{SO}(N)$ groups). 
    \label{tab:modelclasses}}
\end{table}

\subsection{Pseudo-Nambu-Goldstone boson spectrum}\label{sec:pNGBspectrum}

All the models in classes C1-C3  share a common EW structure arising from the $\psi$ sector, as such hyperfermions transform in a real representation of the HC gauge group and are present in five copies. This implies a global flavour symmetry $G_F^\psi = \SU(5)$ which is spontaneously broken by the formation of a condensate $\langle \psi \psi \rangle$. For fermions in a real representation, the condensate is symmetric in the flavour space, which leads to the symmetry breaking pattern~\cite{Dugan:1984hq,Ferretti:2014qta,Agugliaro:2018vsu}
\begin{align}
    \SU(5) \to \SO(5)\,.
\end{align}
This coset yields 14 pNGBs transforming as the $\mathbf{14}$ of $\SO(5)$. Hence, the EW gauge group $\SU(2)_L \times \U(1)_Y$ is embedded within the custodial symmetry $\SU(2)_L \times \SU(2)_R$, itself contained in the unbroken $\SO(5)$ to ensure custodial symmetry protection. The pNGB spectrum can therefore be decomposed under $\SU(2)_L \times \SU(2)_R$ as
\begin{align}
    \mathbf{14} \to (\mathbf{2},\mathbf{2}) \oplus (\mathbf{3},\mathbf{3}) \oplus (\mathbf{1},\mathbf{1})\,.
\end{align}
Here, the $(\mathbf{2},\mathbf{2})$ multiplet is identified with the SM Higgs bi-doublet and contains the three would-be Goldstone bosons eaten by the $W$ and $Z$ bosons together with the physical Higgs boson $h$. The $(\mathbf{3},\mathbf{3})$ bi-triplet further decomposes under the custodial diagonal $\SU(2)_V$ symmetry into a singlet, a triplet and a quintuplet, while the $(\mathbf{1},\mathbf{1})$ corresponds to a neutral scalar singlet. The physical spectrum therefore includes doubly- and singly-charged as well as neutral scalar states, such as the $S^{++}$, $S^+$ and $S^0$ fields listed in \cref{tab:modelclasses}. The properties and LHC phenomenology of these electroweak pNGBs have been studied in detail in~\cite{Cacciapaglia:2022bax}, to which we refer for further discussion.
In addition, an axion-like light pseudo-scalar state arises from the spontaneous breaking of the non-anomalous $\U(1)$ symmetry, and is thus independent of the $\SU(5)/\SO(5)$ coset. 

We now turn to the coloured sector arising from the $\chi$ hyperfermions, which always comprises 6 states: one colour triplet $\chi_{-1/3}$ and a conjugate anti-triplet $\tilde{\chi}_{1/3}$, carrying the appropriate conjugate hypercharges. However, the global flavour symmetry $G_F^\chi$ and its spontaneous breaking pattern depend on whether the $\chi$ states transform in a real, pseudo-real or complex representation of $G_{\mathrm{HC}}$, yielding:
\begin{equation}\label{eq:gfbreaking}\begin{split}
    \text{C1}:& \qquad \SU(6) \to \SO(6) \supset \SU(3)_c \times \U(1)_X\,,\\
    \text{C2}:& \qquad \SU(6) \to \Sp(6) \supset \SU(3)_c \times \U(1)_X\,, \\
    \text{C3}:& \qquad \SU(3)_\chi \times \SU(3)_{\tilde\chi} \times \U(1)_X \to \SU(3)_c \times \U(1)_X\,.
\end{split}\end{equation}
The anomaly-free $\U(1)_X$ symmetry is used to properly define the hypercharge in the composite sector so that $Y = T_R^3 + X$, where $T_R^3$ is the third component of the $\SU(2)_R$ isospin as embedded in $G_F^\psi = \SU(5)_\psi$. In all models of interest, $X = -1/3$, ensuring correct quantum numbers for the top-partner states~\cite{Ferretti:2016upr}. Note that in models of classes C1 and C2, the generator of the $\U(1)_X$ symmetry is directly embedded in $\SU(6)$, while in 
models of class C3 it emerges as the relative phase between the colour triplet and anti-triplet. Finally, the $\chi$ hyperfermions also carry baryon number charges $B = X/2$ acquired via the linear mixing with the third generation quarks: this baryon charge will be carried over to the coloured pNGBs and the hyperbaryons. 

In models of class C1, the six $\chi$ hyperfermions are in a real representation of $G_{\rm HC}$, leading to a global $\SU(6)$ flavour symmetry broken by a symmetric condensate. This produces 20 pNGBs transforming as the $\mathbf{20}$ of $\SO(6)$, which decomposes under $\SU(3)_c \times \U(1)_X$ as\footnote{$\SO(6)$ has three distinct 20-dimensional irreducible representations. In the following, we denote by $\mathbf{20}$ the representation with Dynkin label $(020)$, commonly referred to as $\mathbf{20}'$ in the literature~\cite{Fonseca:2020vke}.}
\begin{align}
    \mathbf{20} \to \mathbf{8}_0 \oplus \mathbf{6}_{-2/3} \oplus \overline{\mathbf{6}}_{ 2/3}\,.
\end{align}
This corresponds to a neutral colour-octet $\pi_8$ and a complex colour-sextet $\pi_6$ state~\cite{Cacciapaglia:2015eqa}. In models of class C2, the $\chi$ hyperfermions lie in a pseudo-real representation so that the $\SU(6)$ flavour symmetry is broken to $\Sp(6)$, producing 14 pNGBs. These decompose as
\begin{align}
    \mathbf{14} \to \mathbf{8}_0 \oplus \mathbf{3}_{2/3} \oplus \overline{\mathbf{3}}_{-2/3}\,,
\end{align}
giving a neutral octet $\pi_8$ and a complex colour triplet $\pi_3$~\cite{Cacciapaglia:2021uqh}. Finally, in models of class C3, the $\chi$ hyperfermions transform in a complex representation of $G_{\rm HC}$, so that the global symmetry $\SU(3)_\chi \times \SU(3)_{\tilde\chi}$ is spontaneously broken to the diagonal $\SU(3)_c$ group. This produces $8$ pNGBs forming a neutral colour-octet $\pi_8$. Unlike for models of classes C1 and C2, there are no additional coloured triplets or sextets in the pNGB spectrum. The colour-octet state $\pi_8$ is therefore a universal feature across all classes of models and can be considered a `standard candle' of partial compositeness at the LHC~\cite{Belyaev:2016ftv}. The full spectrum of coloured pNGBs for the different classes of models, including the triplet and sextet states where present, is summarised in \cref{tab:modelclasses}.

\subsection{Hyperbaryon spectrum}\label{sec:baryonsspectrum}
In addition to the scalar sector, all three classes of models feature a spectrum of fermionic hyperbaryons formed from bound states of $\psi$ and $\chi$ hyperfermions and playing a central role in the realisation of the partial compositeness mechanism. In the class of scenarios considered here, the relevant operators are of the form $\chi \psi \chi$, leading to coloured fermionic resonances transforming in various representations of $\SU(3)_c$ and including states with the appropriate quantum numbers to mix with the elementary top quark. This results in a rich and distinctive phenomenology at hadron colliders, as studied in \cref{sec:pheno}.

In models of classes C1 and C2, the $\chi$ hyperfermions transform in real and pseudo-real representations of the hypercolour group, respectively, and they are related in both cases to a global $\SU(6)$ flavour symmetry. It is possible to construct two types of baryonic operators, yielding the following representations under $\SU(6)$,
\begin{equation}
    \chi \psi \chi \to \mathbf{6} \otimes \mathbf{6} = \mathbf{15}_A \oplus \mathbf{21}_S\qquad \text{and} \qquad
    \bar{\chi} \bar{\psi} \chi \to \mathbf{\bar{6}} \otimes \mathbf{6} = \mathbf{35}_{Adj} \oplus \mathbf{1}\,.
\end{equation}
Under the symmetry breaking $\SU(6) \to \SO(6)$ (class C1), the symmetric product decomposes as $\mathbf{21}_S \to \mathbf{20}_S \oplus \mathbf{1}$, while the antisymmetric product remains irreducible as $\mathbf{15}_A$. Under $\SU(6) \to \Sp(6)$ (class C2) instead, the antisymmetric product splits as $\mathbf{15}_A \to \mathbf{14}_A \oplus \mathbf{1}$, while the symmetric $\mathbf{21}_S$ remains irreducible. The adjoint representation always decomposes as a symmetric plus an antisymmetric representation of the unbroken group, ${\bf 35}_{Adj} \to {\bf 20}_S \oplus {\bf 15}_A$ (for C1) or ${\bf 35}_{Adj} \to {\bf 21}_S \oplus {\bf 14}_A$ (for C2). Hence, in all cases we obtain three independent sets of hyperbaryons, transforming as the following irreducible representations of the unbroken group,
\begin{align} \label{eq:baryon_rep}
    \mathcal{B}_S \equiv \begin{cases} \mathbf{20}_S & \text{(C1)} \\ 
    \mathbf{21}_S & \text{(C2)} \end{cases}\,, \qquad
    \mathcal{B}_A \equiv \begin{cases} \mathbf{15}_A & \text{(C1)} \\ 
    \mathbf{14}_A & \text{(C2)} \end{cases}\,, \qquad
    \mathcal{B}_1 \equiv \mathbf{1}\,,
\end{align}
corresponding to the symmetric representation, the antisymmetric representation and the singlet one respectively, with the singlets arising from the symmetric (antisymmetric) combination in class C1 (C2) and the singlet operator. The masses of the states within each multiplet are approximately degenerate, up to corrections from the SM gauge interactions and the mixing with the top quark, while $\mathcal{B}_S$, $\mathcal{B}_A$ and $\mathcal{B}_1$ can in general have independent masses. Under $\SU(3)_c \times \U(1)_X$, we obtain the following decompositions 
\begin{align}\label{eq:baryons_C1C2}
    \mathcal{B}_S &\to \begin{cases} 
        \mathbf{8}_0 \oplus \mathbf{6}_{-2/3} \oplus \overline{\mathbf{6}}_{2/3} 
        & \text{(C1)}\,, \\[4pt]
        \mathbf{8}_0 \oplus \mathbf{6}_{-2/3} \oplus \overline{\mathbf{6}}_{2/3} 
        \oplus \mathbf{1}_0 & \text{(C2)}\,,
    \end{cases}\quad 
    \mathcal{B}_A \to \begin{cases} 
        \mathbf{8}_0 \oplus \mathbf{3}_{2/3} \oplus \overline{\mathbf{3}}_{-2/3} 
        \oplus \mathbf{1}_0 & \text{(C1)}\,, \\[4pt]
        \mathbf{8}_0 \oplus \mathbf{3}_{2/3} \oplus \overline{\mathbf{3}}_{-2/3} 
        & \text{(C2)}\,,
    \end{cases}
\end{align}
while the singlet $\mathcal{B}_1$ has zero $X$-charge. All hyperbaryons also transform as a $\bf 5$ of the unbroken $\SO(5)$ in the $\psi$ sector, which fixes the electroweak charges of the various components (see below). The spectrum thus contains fermionic resonances in all colour representations appearing in the pNGB sector, together with additional states. In particular, the symmetric combination $\mathcal{B}_S$ gives rise to colour-octet states $\mathcal Q_8$ and colour-sextet states $\mathcal Q_6$ and $\mathcal Q_6^c$, while the antisymmetric combination $\mathcal{B}_A$ contains colour octets $Q_8$ and colour triplets $Q_3$ and $Q_3^c$. The colour-triplet states carry precisely the quantum numbers required to act as top partners and are therefore the states that mix with the elementary $q_L$ and $t_R^c$ fields through partial compositeness.  In addition, two independent colour-singlet multiplets, $Q_1$ and $\hat Q_1$, are present, which we name such that $\hat Q_1$ originates from the $\mathcal{B}_1$ multiplet. Notably, colour-octet resonances arise in the two $\mathcal{B}_S$ and $\mathcal{B}_A$ combinations, leading to two distinct octet multiplets with potentially different masses and phenomenology: that of the octet partner of the triplets, which couples directly to the top quark, has been studied in~\cite{Cacciapaglia:2021uqh}. Finally, the two colour-singlet multiplets give rise to charged and neutral Dirac and Majorana fermions that can play the role of a dark matter candidate. The detailed field content of each multiplet follows from the $\SO(5)$ five-plet structure associated with the $\psi$ component of the baryons, as discussed below.

In constructions of class C3, the $\chi$ hyperfermions transform in a complex representation of the hypercolour gauge group $G_{\rm HC}$, thereby coming in conjugate pairs $\chi \in R$ and $\tilde{\chi} \in \overline{R}$ with $R=\mathbf{F}$ for the model M6 (with $G_{\rm HC} = \SU(4)$) and $R=\mathbf{Spin}$ for the model M7 (with $G_{\rm HC} = \SO(10)$). The global flavour symmetry of the coloured sector is therefore $\SU(3)_\chi \times \SU(3)_{\tilde{\chi}} \times \U(1)_X$, under which the hyperfermions transform as
\begin{align}
    \chi \in (\mathbf{3}, \mathbf{1})_{-1/3}\,, \qquad \tilde{\chi} \in (\mathbf{1}, \overline{\mathbf{3}})_{1/3}\,.
\end{align}
The determination of the set of allowed hypercolour-singlet baryonic operators requires combining hyperfermions in representations of $G_{\rm HC}$ such that the total product contains a hypercolour singlet. This allows several classes of operators,
\begin{align}\label{eq:C3_allowedbaryons}
    \chi\psi\chi\,, \qquad 
    \tilde{\chi}\psi\tilde{\chi}\,, \qquad
    \bar{\chi}\bar{\psi}\tilde{\chi}\,, \qquad
    \chi\psi\bar{\tilde{\chi}}\,,
\end{align}
where the last two operators involve the conjugate hyperfermions $\bar{\chi}$, $\bar{\tilde{\chi}}$ and $\bar{\psi}$, and whereas operators of the type $\chi \psi \tilde{\chi}$ are forbidden by gauge invariance. In both models M6 and M7, the representations of $\chi$ and $\psi$ listed in \cref{tab:modelclasses} ensure that all four allowed combinations contain a hypercolour singlet, and can therefore contribute to the baryonic spectrum. The four operators of \cref{eq:C3_allowedbaryons} can first be decomposed under the two flavour groups $\SU(3)_\chi \times \SU(3)_{\tilde{\chi}}$ as
\begin{equation}\begin{split}
    &\chi\psi\chi \in (\mathbf{3} \otimes \mathbf{3}, \mathbf{1}\otimes \mathbf{1}) = (\mathbf{6}, \mathbf{1}) \oplus (\overline{\mathbf{3}}, \mathbf{1})\,, \hspace{1cm}
 \chi\bar{\psi}\bar{\tilde{\chi}} \in (\mathbf{3}\otimes \mathbf{1}, \mathbf{1}\otimes \mathbf{3}) = (\mathbf{3}, \mathbf{3})\,,\\[.2cm]
    & \bar{\chi}\bar{\psi}\tilde{\chi} \in (\overline{\mathbf{3}}\otimes \mathbf{1}, \mathbf{1}\otimes \overline{\mathbf{3}}) = (\overline{\mathbf{3}}, \overline{\mathbf{3}})\,, \hspace{2.3cm}\!
        \tilde{\chi}\psi\tilde{\chi} \in (\mathbf{1}\otimes \mathbf{1}, \overline{\mathbf{3}} \otimes \overline{\mathbf{3}}) = (\mathbf{1}, \overline{\mathbf{6}}) \oplus (\mathbf{1}, \mathbf{3})\,,       
\end{split}\end{equation}
where the operators in the top line carry an $X$-charge of $-2/3$, while those on the bottom line carry an $X$-charge of $2/3$.
In a second step, we project these flavour representations onto the diagonal QCD colour group $\SU(3)_c$ and assign the corresponding $U(1)_X$ charges. We emphasise that the $X$-charge of the conjugate operators coincides with that of the original ones, because $\bar{\chi}$ and $\bar{\tilde{\chi}}$ carry opposite $X$-charge while $\psi$ and $\bar{\psi}$ are neutral. This yields the following baryon multiplets
\begin{equation}\label{eq:C3baryons}\begin{split}
    \chi\psi\chi &: 
       \hspace{.3cm} (\mathbf{6},\mathbf{1})_{-2/3} \to \mathbf{6}_{-2/3} \equiv \mathcal{Q}_6, 
       \hspace{.5cm} (\overline{\mathbf{3}},\mathbf{1})_{-2/3} \to \overline{\mathbf{3}}_{-2/3} \equiv Q_3^c\,,\\[1mm]
    \tilde{\chi}\psi\tilde{\chi} &: 
       \hspace{.3cm} (\mathbf{1},\overline{\mathbf{6}})_{2/3} \to \overline{\mathbf{6}}_{2/3} \equiv \mathcal{Q}_6^c, 
       \hspace{.7cm} (\mathbf{1},\mathbf{3})_{2/3} \to \mathbf{3}_{2/3} \equiv Q_3\,,\\[1mm]
    \bar{\chi}\bar{\psi}\tilde{\chi} &:
       \hspace{.3cm} (\overline{\mathbf{3}},\overline{\mathbf{3}})_{2/3} \to \overline{\mathbf{6}}_{2/3} \oplus \mathbf{3}_{2/3} \equiv \mathcal{Q}_6' \oplus {Q_3^c}'\,,\\[1mm]
    \chi\bar{\psi}\bar{\tilde{\chi}} &: 
       \hspace{.3cm} (\mathbf{3},\mathbf{3})_{-2/3} \to \mathbf{6}_{-2/3} \oplus \overline{\mathbf{3}}_{-2/3} \equiv {\mathcal{Q}_6^c}' \oplus Q_3'\,.
\end{split}\end{equation}
Each of the hyperbaryons above have, in principle, an independent mass. A distinctive feature of models of class C3 is the absence of colour-octet and colour-singlet fermionic resonances, in stark contrast to scenarios of classes C1 and C2. This follows directly from the complex representation structure: the four possible operators involve only one type of representation (either two $\mathbf{3}$ or two $\overline{\mathbf{3}}$, and never a $\mathbf{3} \otimes \overline{\mathbf{3}}$ combination), so their products under $\SU(3)_c$ can only yield sextets and triplets, and never an adjoint or a singlet. 

The partial compositeness couplings require the composite baryonic operators to mix with the elementary SM fields through spurions transforming in the fundamental representation of $\SO(5)$. Under the decomposition $\SO(5) \supset \SU(2)_L \times \SU(2)_R$, this 
five-plet splits as
\begin{align}\label{eq:fiveplet}
    \mathbf{5} \to (\mathbf{2}, \mathbf{2}) \oplus (\mathbf{1}, \mathbf{1}),
\end{align}
where the bi-doublet $(\mathbf{2},\mathbf{2})$ contains states that can mix with the SM left-handed quark doublet $q_L$ and the singlet $(\mathbf{1},\mathbf{1})$ contains states that can mix with the right-handed top quark $t_R^c$. The different hyperbaryon states inherit their EW charges from the combination of the $\SO(5)$ five-plet and the $X$-charge of the constituent hyperfermions, and the hypercharge is given by $Y = T^3_R + X$. The decomposition under the SM gauge group $G_\text{SM} = \SU(3)_c \times \SU(2)_L \times \U(1)_Y$ yields, for the sextet states $\mathcal{Q}_6$ 
\begin{equation}\begin{split}
    \mathcal Q_6: \quad (\mathbf 6, \mathbf 2)_{- 1/6} \oplus (\mathbf 6, \mathbf 2)_{- 7/6} \oplus (\mathbf 6, \mathbf 1)_{- 2/3} &\quad \text{of} \,\, G_\text{SM}\,, \\
    \hspace{1cm} \to \mathbf 6_{-5/3 } \oplus (3\times \mathbf 6_{- 2/3}) \oplus \mathbf 6_{ 1/3} & \quad \text{of} \,\, \SU(3)_c\times \U(1)_Q\,,
\end{split}\end{equation}
where the subscript in the second line refers to the electric charge which will be used, in the following, to denote the mass eigenstates as $\mathcal Q_6^{-5/3}, \mathcal Q_6^{-2/3}$ and $\mathcal Q_6^{1/3}$. Here, the three states with charge $-2/3$ arise from both the bi-doublet and the singlet components of the five-plet, and in principle they mix.

Analogously, the colour-triplet hyperbaryons, denoted $Q_3$, form five-plets of $\SO(5)$, which decompose under $\SO(5) \supset \SU(2)_L \times \SU(2)_R$ as given in \cref{eq:fiveplet}. As already mentioned, the bi-doublet $(\mathbf{2},\mathbf{2})$ contains states that can mix with the SM left-handed quark doublet $q_L$, while the singlet $(\mathbf{1},\mathbf{1})$ mixes with the right-handed top quark $t_R^c$. To see this in practice, we combine this $\SO(5)$ structure with the $X$-charge of the constituent hyperfermions. The colour-triplet states then decompose under the SM gauge group as
\begin{equation}\label{eq:tripletQN}\begin{split}
    Q_3: \quad (\mathbf{3}, \mathbf{2})_{1/6} \oplus (\mathbf{3}, \mathbf{2})_{7/6} \oplus (\mathbf{3}, \mathbf{1})_{2/3} & \quad \text{under } G_\text{SM}\,, \\
    \hspace{1cm} \to \mathbf{3}_{5/3} \oplus (3 \times \mathbf{3}_{2/3}) \oplus \mathbf{3}_{-1/3} & \quad \text{under} \,\, \SU(3)_c\times \U(1)_Q\,.
\end{split}\end{equation}
Using conventional notation, we denote the mass eigenstates by their electric charges as $X_{5/3}$, $X_{2/3}$, $T_L$, $B_L$ and $T_R$, where $Q_L = (T_L, B_L)^T$ is the composite doublet that mixes with $q_L$ and corresponds to the $(\mathbf{3}, \mathbf{2})_{1/6}$ state of $G_\text{SM}$ in \cref{eq:tripletQN}, $T_R$ mixes with $t_R^c$ and is the $(\mathbf{3}, \mathbf{1})_{2/3}$ of $G_\text{SM}$ appearing in \cref{eq:tripletQN}, and $X = (X_{5/3}, X_{2/3})^T$ is an exotic doublet with electric charges $5/3$ and $2/3$ corresponding to $(\mathbf{3}, \mathbf{2})_{7/6}$ in \cref{eq:tripletQN}. This hence yields the standard naming scheme definition
\begin{align} \label{eq:nomenclature}
    Q_3 = (X, \quad Q_L,\quad  iT_R) \equiv \mqty( \mqty( X_{5/3} \\[.1cm] X_{2/3} ),\quad  \mqty(T_L \\[.1cm] B_L),  \quad iT_R )\,.
\end{align}
It should be noted that while all the states with electric charge of $2/3$ can mix, the state $X_{2/3}$ only mixes weakly with the top quark since two Higgs insertions are required to match the hypercharges.

Furthermore, the colour-octet baryons $\mathcal{Q}_8 \subset \mathcal{B}_S$ and $Q_8 \subset \mathcal{B}_A$ appearing in models of classes C1 and C2 both carry vanishing $X$-charge, so that their hypercharge is entirely determined by their $\SU(2)_R$ isospin quantum numbers ($Y = T_R^3$). Combining with the $\SO(5)$ five-plet structure of \cref{eq:fiveplet}, their decomposition under the SM gauge group yields
\begin{equation}\begin{split}
    \mathcal{Q}_8,\, Q_8: \quad 
    (\mathbf{8}, \mathbf{2})_{1/2} \oplus (\mathbf{8}, \mathbf{2})_{-1/2} 
    \oplus (\mathbf{8}, \mathbf{1})_{0} & \quad \text{under } G_\text{SM}\,, \\
    \to \mathbf{8}_{1} \oplus (3\times\mathbf{8}_{0}) \oplus \mathbf{8}_{-1} \hspace{1.4cm}
    & \quad \text{under } \SU(3)_c \times \U(1)_Q\,,
\end{split}\end{equation}
the two octet multiplets thus giving rise to pairs of electrically-charged states $\mathcal{Q}_8^{\pm 1}$ and $Q_8^{\pm 1}$ and two sets of three neutral states $\mathcal{Q}_8^0$ and $Q_8^0$, with one per multiplet being a Majorana state. The latter originate from two independent neutral components of the bi-doublet $(\mathbf{2},\mathbf{2})$ and one from the singlet $(\mathbf{1},\mathbf{1})$ of the $\SO(5)$ five-plet. The two octet multiplets $\mathcal{Q}_8$ and $Q_8$ are in general non-degenerate in mass, as they originate from different representations $\mathcal{B}_S$ and $\mathcal{B}_A$. Models of classes C1 and C2 also include the colour-singlet baryons $Q_1$ and  $\hat{Q}_1$, and both carry vanishing $X$-charge and colour charge. Their hypercharge is therefore entirely determined by their $T_R^3$ quantum numbers so that their embedding into $\SO(5)$ five-plets yields the decomposition
\begin{equation}\begin{split}
    Q_1,\, \hat{Q}_1: \quad 
    (\mathbf{1}, \mathbf{2})_{1/2} \oplus (\mathbf{1}, \mathbf{2})_{-1/2} 
    \oplus (\mathbf{1}, \mathbf{1})_{0} & \quad \text{under } G_\text{SM}\,, \\
    \to \mathbf{1}_{1} \oplus (3\times\mathbf{1}_{0}) \oplus \mathbf{1}_{-1} \hspace{1.4cm}
    & \quad \text{under } \SU(3)_c \times \U(1)_Q\,.
\end{split}\end{equation}
Each of them thus contains Weyl fermions that combine into a singly charged Dirac fermion $Q_1^\pm$ and an electrically neutral Dirac fermion $Q_1^0$, originating from the bi-doublet, as well as one neutral Majorana fermion arising from the singlet component of the five-plet. The two Majorana states, denoted $\tilde{q}_1^0$ and $\hat{\tilde{q}}_1^0$, are uncharged under all SM gauge interactions and therefore belong to real representations, allowing for Majorana mass terms and highlighting their qualitative difference from the other neutral states~\cite{Cacciapaglia:2021uqh}. In addition, states within a given multiplet are approximately degenerate in mass, up to SM gauge corrections, while the two multiplets $Q_1$ and $\hat{Q}_1$ can, in general, have different masses. Finally, the neutral states may provide viable dark matter candidates.

\subsection{Interactions of the colour-sextet hyperbaryons}\label{sec:sextets}
In this section, we turn to the interactions that govern the decays of the colour-sextet states $\mathcal{Q}_6$ which are the primary focus of this work. Two distinct sources of interactions contribute to the phenomenologically relevant couplings: the derivative couplings of the hyperbaryons to the pNGBs, and additional couplings induced by the partial compositeness mechanism. We discuss each in turn.

The leading interactions between the hyperbaryons and the pNGBs are described within the CCWZ Lagrangian~\cite{Coleman:1969sm, Callan:1969sn} by a term that, at the lowest order in the pNGB fields, takes the generic form
\begin{align}\label{eq:lder}
    \mathcal{L}_{\rm der} = \sum_{i,j}\bigg[\frac{c_{ij}}{f_\chi}\, 
    \bar{\mathcal{B}}_i\, \bar\sigma^\mu \partial_\mu \Pi_\chi\, 
    \mathcal{B}_j + \mathrm{H.c.}\bigg]\,,
\end{align}
where $\Pi_\chi$ is the pNGB matrix of the coloured sector, $f_\chi$ is the associated decay constant and $c_{ij}$ are $\mathcal{O}(1)$ dimensionless coefficients encoding the underlying strong dynamics. Physically, this operator describes the emission of a pNGB in transitions between composite baryons belonging to different representations of the unbroken group, and the symmetry structure of the strong sector constrains which combinations of baryons can couple to one another. In particular, a non-vanishing interaction requires the product $\bar{\mathcal{B}}_i \otimes \mathcal{B}_j$ to contain the representation of the unbroken group in which the pNGBs transform. As a consequence, not all pairs $(\mathcal{B}_i, \mathcal{B}_j)$ are allowed to couple. For our purposes, the most relevant couplings involve hyperbaryons in different representation, one of which containing the sextet. The rationale behind this is the fact that components within the sextet multiplet are near degenerate, hence decays within the multiplet will be highly suppressed. Furthermore, as all hyperbaryons are in the same representation of the EW $\SO(5)$ group, the relative indices can be trivially traced without any insertion of the corresponding pNGBs in \cref{eq:lder}. As a consequence, the couplings between the hyperbaryons and the coloured pNGBs generated by \cref{eq:lder} are unsuppressed by the Higgs vacuum expectation value $v$.\footnote{More generally, operators involving the electroweak pNGBs will be present. However, they can only mediate decays within the same colour multiplet, which are highly suppressed by the mass degeneracy.}

A coupling common to all model classes C1, C2 and C3 is the $\mathcal{Q}_6$-$Q_3$-$\pi_8$ vertex. It originates from the derivative interactions in \cref{eq:lder} once the pNGB matrix $\Pi_\chi$ is expanded in terms of the constituent coloured states, in particular the colour-octet pNGB $\pi_8 \sim \mathbf{8}$ of $\SU(3)_c$. Independently of the details of the underlying coset, the existence of this coupling is guaranteed by the QCD colour symmetry, since the product of the involved representations satisfies 
\begin{align}
    \mathbf{3} \otimes \mathbf{6} \otimes \mathbf{8} \supset \mathbf{1}\,.
\end{align}
The corresponding interaction Lagrangian can be written as
\begin{align}\label{eq:Lag368}
    \mathcal{L}_{\rm der} \supset \frac{c}{f_\chi} \left(
      -i\, \bar{Q}_3^{c,i}\, \bar\sigma^\mu\, \partial_\mu\pi_8^a\, \mathcal{Q}_6^s\, J_{sia}
      -i\, \bar{Q}_{3,i}\, \bar\sigma^\mu\, \partial_\mu\pi_8^a\, \mathcal{Q}_{6,s}^c\, \bar{J}^{sia} 
      + \mathrm{H.c.} \right)\,,
\end{align}
where $J_{sia}$ and $\bar{J}^{sia}$ are the Clebsch-Gordan coefficients for $\mathbf{3} \otimes \mathbf{6} \otimes \mathbf{8}$ in the conventions of~\cite{Carpenter:2021rkl}, and  $s$, $i$ and $a$ denote colour indices in the sextet, fundamental and adjoint representation respectively. Since some of the components of the $Q_3$ multiplet mix with the top fields $q_L$ and $t_R^c$ with large mixing angles, they can generate direct and large couplings to the top mass eigenstates via the octet pNGB.

In models of class C1, the presence of the colour-sextet pNGB $\pi_6$ introduces additional decay channels for the sextet baryons, and similarly in constructions from class C2 where the colour-triplet $\pi_3$ mediates analogous interactions. The derivative interactions of \cref{eq:lder} dictate the structure of these couplings, which can be written as 
\begin{equation}\label{eq:CCWZ_C1C2}\begin{split}
  \mathcal{L}_{\rm der}^{\rm C1} &\supset 
    \frac{c}{f_\chi} \, \bar{Q}_1\, \bar\sigma^\mu \partial_\mu \pi_6^s \, \mathcal{Q}^c_{6,s} 
    - \frac{c}{f_\chi} \, \bar{Q}_1\, \bar\sigma^\mu \partial_\mu \pi_{6,s}^c \, \mathcal{Q}_6^s 
    - \frac{c}{\sqrt{2}f_\chi} \bar{Q}_8^a \, \bar\sigma^\mu \partial_\mu \pi_{6,s'}^c \, \mathcal{Q}_6^s \, [t_6^a]^{s'}{}_{\!s}\\
    &\qquad+ \frac{c}{\sqrt{2}f_\chi} \bar{Q}_8^a \, \bar\sigma^\mu \partial_\mu \pi_6^s \, \mathcal{Q}_{6,s'}^c \, [t_6^a]^{s'}{}_{\!s}
    + \frac{\hat{c}}{f_\chi} \, \bar{\hat{Q}}_1 \, \bar\sigma^\mu \partial_\mu \pi_6^s \, \mathcal{Q}_{6,s}^c
    + \frac{\hat{c}}{f_\chi} \, \bar{\hat{Q}}_1 \, \bar\sigma^\mu \partial_\mu \pi_{6,s}^c \, \mathcal{Q}_6^s 
    + \mathrm{H.c.}\,,\\
  \mathcal{L}_{\rm der}^{\rm C2} &\supset 
    \frac{i c}{f_\chi} \, \bar{Q}_8^a\, \bar\sigma^\mu \partial_\mu \pi_{3,i}^c \, \mathcal{Q}_{6,s}^c \, \bar{J}^{sia} 
    + \frac{i c}{f_\chi} \, \bar{Q}_8^a\, \bar\sigma^\mu \partial_\mu \pi_3^i \, \mathcal{Q}_6^s \, J_{sia} 
    + \mathrm{H.c.}\,,
\end{split}\end{equation}
where the matrices $t_6$ stand for the generators of $\SU(3)$ in the sextet representation. The structure of \cref{eq:CCWZ_C1C2} reflects the symmetry properties of the strong sector. In models of class C1, the sextet baryon $\mathcal{Q}_6$ couples to both the singlet and octet baryons $Q_1$ and $Q_8$ with a common coefficient $c$, as these interactions originate from the same CCWZ operator as the $\mathcal{Q}_6$-$Q_3$-$\pi_8$ vertex in \cref{eq:Lag368}. They thus correspond to different components of a single invariant of the unbroken symmetry. In addition, the singlet $\hat{Q}_1$, which belongs to a different representation, couples with an independent coefficient $\hat{c}$. In models of class C2, the sextet baryon instead couples to the octet baryon $Q_8$ through the emission of the colour-triplet pNGB $\pi_3$, with a structure fixed by $\SU(3)_c$ invariance. In contrast, models of class C3 do not feature additional coloured pNGBs beyond the octet $\pi_8$, and therefore do not generate any analogous interactions. The complete derivation of these expressions is collected in Appendix~\ref{app:calculation}.

The second source of couplings arises from the linear mixing of elementary SM quarks with composite operators via partial compositeness. The relevant Lagrangian takes the generic form 
\begin{align}\label{eq:PC_general}
    \mathcal{L}_{\rm PC} \supset 
    -\lambda_L\, {q}_L\, \mathcal{O}_L 
    - \lambda_R\, {t}_R^c\, \mathcal{O}_R + \mathrm{H.c.}\,,
\end{align}
where $\mathcal{O}_{L,R}$ are composite operators constructed in terms of the hyperfermions $\psi$ and $\chi$ and containing components with the quantum numbers of the SM fields $q_L$ and $t^c_R$ (\textit{i.e.}\ colour triplets and anti-triplets). In the low-energy Lagrangian, these interactions then generate couplings with all hyperbaryons that have a non-zero overlap with $\mathcal{O}_L$ and/or $\mathcal{O}_R$. We have seen in the previous subsections that the sextet and triplet states can be generated by the same operators: for models of classes C1 and C2, this occurs as long as $\mathcal{O}_{L/R}$ transform as the adjoint representation of $\SU(6)_\chi$, while for models of class C3 it occurs instead if the operators transform as a bi-fundamental of the $\SU(3)_\chi \times \SU(3)_{\tilde\chi}$ global symmetry. The low-energy Lagrangian subsequently contains couplings of the form
\begin{align}\label{eq:PC_sextet}
    \mathcal{L}_{\rm PC} \supset 
    \tilde{\lambda}_L\, \zeta_L\, \pi_8\, \mathcal{Q}_6 + \tilde{\lambda}_R\, \zeta_R^c\, \pi_8\, \mathcal{Q}_6^c
     + \mathrm{H.c.}\,,
\end{align}
with $\zeta_L$ and $\zeta^c_R$ denoting incomplete $\SU(5)$ representations containing the SM $q_L$ and $t^c_R$ states, respectively, as detailed in Appendix~\ref{app:pcint}. Moreover, the parameters $\tilde{\lambda}_{L/R}$ stand for some effective couplings proportional to $\lambda_{L/R}$, and the appropriate components of the sextets will be selected by the spurion structure in the $\SU(5)$ space (which we leave understood in the above expression). We remark that the same structures are relevant for the linear mixing of the triplets $Q_3$ with the SM top fields, hence the couplings above only give additional contributions to decay channels already generated by the interactions included in \cref{eq:Lag368}. A more detailed discussion will be presented in the next section.

\subsection{Exotically charged baryons: the model M12}\label{sec:M12}

The model M12, classified as class C4 in \cref{tab:modelclasses}, is notable for allowing hyperbaryons of both types, $\psi\chi\psi$ and $\chi\psi\chi$. Its hypercolour group is $G_\mathrm{HC} = \SU(5)$, and the representations of the hyperfermions are $(\psi, \tilde \psi) \in (\mathbf F, \mathbf{\bar F})$ and $(\chi, \tilde \chi) \in (\mathbf A_2, \overline{\mathbf A}_2)$. The global symmetry is therefore $G_F = \SU(4)_\psi \times \SU(4)_{\tilde{\psi}} \times \SU(3)_\chi \times \SU(3)_{\tilde{\chi}} \times \U(1)^3$. Using the properties of the $\SU(5)$ representations, one may construct hypercolour singlets from the tensor products
\begin{align}
    \mathbf 5 \otimes \mathbf{10} \otimes \mathbf{10}
    \qquad\text{and}\qquad
    \overline{\mathbf{10}} \otimes \mathbf 5 \otimes \mathbf 5\,,
\end{align}
and from their conjugate structures. These give rise to the composite operators such as
\begin{align}
    \chi \psi \chi\,, \quad \chi \bar{\tilde \psi} \bar{\tilde \chi} \qquad \text{and}\qquad
    \psi \tilde \chi \psi\,, \quad \psi \bar \chi \bar{\tilde \psi}
    \,.
\end{align}
Among these states, only the baryons of type $\psi\chi\psi$ can provide viable top-partner candidates, and they transform only as colour triplets or anti-triplets so that no exotic QCD representations arise in this case. By contrast, the $\chi\psi\chi$ operators are phenomenologically disfavoured for sourcing the top mass but give rise to a spectrum of additional exotic coloured fermions. The reason is that the electroweak quantum numbers of such $\chi\psi\chi$ baryons are controlled by the single $\psi$ field that they contain. Indeed, under the global symmetry one has
\begin{align} \label{eq:M12psi1}
    \psi \in (\mathbf 4, \mathbf 1), \qquad \tilde \psi \in (\mathbf 1, \mathbf {\bar 4}) \qquad \text{of }\quad  \SU(4)_\psi \times \SU(4)_{\tilde{\psi}},
\end{align}
which decompose under the custodial subgroup $\SU(2)_L \times \SU(2)_R$ as~\cite{Ferretti:2016upr}
\begin{align} \label{eq:M12psi2}
    (\mathbf 4, \mathbf 1) \to (\mathbf 2, \mathbf 1) \oplus (\mathbf 1, \mathbf 2), \qquad (\mathbf 1, \mathbf {\bar 4}) \to (\mathbf 2, \mathbf 1) \oplus (\mathbf 1, \mathbf 2)\,.
\end{align}
Therefore, the baryons of type $\chi\psi\chi$ do not comprise a bidoublet which would be suitable for realising the usual custodially protected partial-compositeness structure, and thus induce large corrections to the $Z$ coupling to left-handed bottom quarks~\cite{Agashe:2006at}.

We now focus on the exotic coloured fermions arising from the $\chi\psi\chi$ operators. In order to determine their EW quantum numbers and decay patterns, we first need to specify the hypercharge embedding. In model M12, the hypercharge is embedded partly in the custodial subgroup of the EW coset and partly in the anomaly-free Abelian symmetries of the underlying hyperfermion theory. Since the $\psi$ condensate breaks $\SU(4)_\psi \times \SU(4)_{\tilde\psi} \to \SU(4)_D \supset \SU(2)_L \times \SU(2)_R$,\footnote{The properties of the $15$ electroweak pNGBs arising from this symmetry-breaking pattern have been studied in~\cite{Ma:2015gra,Wu:2017iji}.} the custodial contribution is provided by the diagonal generator $T_R^3$. In addition, the theory contains two independent vector-like $\U(1)$ charges, which we denote by $X_\psi$ and $X_\chi$. The hypercharge is therefore embedded as
\begin{align}
    Y \equiv T_R^3 + X_\psi + X_\chi\,.
\end{align}
The hyperfermions $\chi$ transform under $\SU(3)_\chi\times\SU(3)_{\tilde\chi}\times\U(1)_\chi$ as
\begin{align}
    \chi \in (\mathbf3,\mathbf1)_{X_\chi}\qquad\text{and}\qquad
    \tilde\chi \in (\mathbf1,\overline{\mathbf3})_{-X_\chi}\,,
\end{align}
which reduce under the diagonal colour subgroup $\SU(3)_c\times\U(1)_\chi$ to
\begin{align}
    \chi \to \mathbf3_{X_\chi}\qquad\text{and}\qquad
    \tilde\chi \to \overline{\mathbf3}_{-X_\chi}\,.
\end{align}
Similarly, for the hyperfermions $\psi$ one has
\begin{equation}\begin{split}
    \psi \in (\mathbf4,\mathbf1)_{X_\psi}
    &\;\to\; \mathbf4_{X_\psi}
    \;\to\; (\mathbf2,\mathbf1)_{X_\psi}\oplus(\mathbf1,\mathbf2)_{X_\psi}\,,\\[.1cm]
    \tilde\psi \in (\mathbf1,\overline{\mathbf4})_{-X_\psi}
    &\;\to\; \overline{\mathbf4}_{-X_\psi}
    \;\to\; (\mathbf2,\mathbf1)_{-X_\psi}\oplus(\mathbf1,\mathbf2)_{-X_\psi}\,,
\end{split}\end{equation}
where the first arrow corresponds to the reduction to $\SU(4)_D\times\U(1)_\psi$ and the second to the decomposition under $\SU(2)_L\times\SU(2)_R\times\U(1)_\psi$. The values of $X_\psi$ and $X_\chi$ are then fixed by requiring that the baryons identified as top partners reproduce the SM hypercharges. For instance, the EW-singlet colour anti-triplet contained in $\psi\tilde\chi\psi$ must have $Y=-2/3$, like the right-handed top quark with which it mixes. This gives
\begin{align}
    Y = 0 + 2X_\psi - X_\chi = -\frac23
    \qquad\Rightarrow\qquad
    X_\chi - 2X_\psi = \frac23\,.
\end{align}
With this choice, the baryons that mix with $q_L$ also acquire the correct hypercharges.

With this information, we can determine the hypercharges of the $\chi\psi\chi$-type hyperbaryons. The relevant operators transform under the non-Abelian part of $G_F$ as
\begin{align}
    \chi \psi \chi \in  ({\bf 4},\, {\bf 1},\, {\bf 6}\oplus {\bf \bar 3}, {\bf 1})_{X_\psi + 2 X_\chi}\quad\text{and}\quad 
    \chi \bar{\tilde\psi} \bar{\tilde\chi} \in  ({\bf 1},\, {\bf 4},\, {\bf 3}, {\bf 3})_{X_\psi + 2 X_\chi}\,, 
\end{align}
while $\tilde{\chi}\tilde{\psi}\tilde{\chi}$ and $\tilde{\chi} \bar{\psi}\bar{\chi}$ transform in the conjugate representations, respectively. Using the relation $X_\chi-2X_\psi=2/3$, the Abelian charge can be written as $X_\psi+2X_\chi = 5X_\psi+4/3$. Here and in the following, the subscript denotes the charge under $Y-T_R^3$, and we display the representations under the non-Abelian part of $G_F$. Under the unbroken group $\SU(4)_D\times\SU(3)_c$, the two operators above transform in the same way. Hence, this class of hyperbaryons contains two multiplets with the same electroweak quantum numbers: one transforming as a sextet and the other as an anti-triplet under QCD. More explicitly, the sextet and anti-triplet baryons decompose under the SM gauge group as 
\begin{equation}\begin{split}
    & \mathcal{B}_6 \in ({\bf 4},\, {\bf 6})_{5X_\psi+4/3} \to ({\bf 6},\, {\bf 2})_{5X_\psi+4/3} \oplus ({\bf 6},\, {\bf 1})_{5X_\psi+11/6} \oplus ({\bf 6},\, {\bf 1})_{5X_\psi+5/6}\,,\\[.2cm]
    & \mathcal{B}_{\bar{3}} \in ({\bf 4},\, {\bf \bar 3})_{5X_\psi+4/3} \to ({\bf \bar 3},\, {\bf 2})_{5X_\psi+4/3} \oplus ({\bf \bar 3},\, {\bf 1})_{5X_\psi+11/6} \oplus ({\bf \bar 3},\, {\bf 1})_{5X_\psi+5/6}\,.
\end{split}\end{equation}
If one of the anti-triplet singlets has the same SM quantum numbers as the $b_R^c$ or $t_R^c$ fields, it can mix with the corresponding elementary quark through partial compositeness, and this mixing can then mediate the decay of the sextet into a SM quark plus the colour-octet pNGB $\pi_8$. This requirement leads to three possible hypercharge assignments,
\begin{equation}\renewcommand{\arraystretch}{1.4}\left\{\begin{array}{l l c l l}
    5 X_\psi + \frac{11}{6} = -\frac{2}{3}\,, \quad  & 5X_\psi + \frac{5}{6} = -\frac{5}{3} \quad 
      &\Rightarrow  \ & X_\psi = -\frac{1}{2}\,,\quad & X_\chi = -\frac{1}{3}\,, \\
    5 X_\psi + \frac{11}{6} = \frac{1}{3}\,, \quad & 5X_\psi + \frac{5}{6} = -\frac{2}{3} 
      & \Rightarrow \ & X_\psi = -\frac{3}{10}\,, \quad & X_\chi = \frac{1}{15}\,, \\
    5 X_\psi + \frac{11}{6} = \frac{4}{3}\,, \quad & 5X_\psi + \frac{5}{6} = \frac{1}{3} \quad 
      & \Rightarrow \ & X_\psi = -\frac{1}{10}\,, \quad & X_\chi = \frac{7}{15} \,.
\end{array}\right.\end{equation}
With the first assignment, one of the two anti-triplet singlets carries precisely the hypercharge of the $t_R^c$ field. After inserting the mixing of this anti-triplet baryon with the corresponding elementary quark field, the sextet components effectively decay into SM third-generation quarks. With the second assignment, two anti-triplets carry the hypercharges of the $q_L$ and $t_R^c$ states, hence their contribution to the top mass must be subleading to avoid a too large contribution to the $Zb_L \bar b_L$ coupling.
Although the resulting sextet decays can lead to final states similar to those generated by the partial-compositeness interactions discussed in \cref{eq:PC_sextet}, the M12 construction differs in its hypercharge assignment and in the structure of the anti-triplet states that mediate the decay. A complete treatment would therefore require a dedicated analysis of the allowed mixings and custodial-symmetry constraints, which lies beyond the scope of the present work. We therefore do not include the model M12 in the phenomenological analysis below.

\section{Phenomenology at the LHC}\label{sec:pheno}

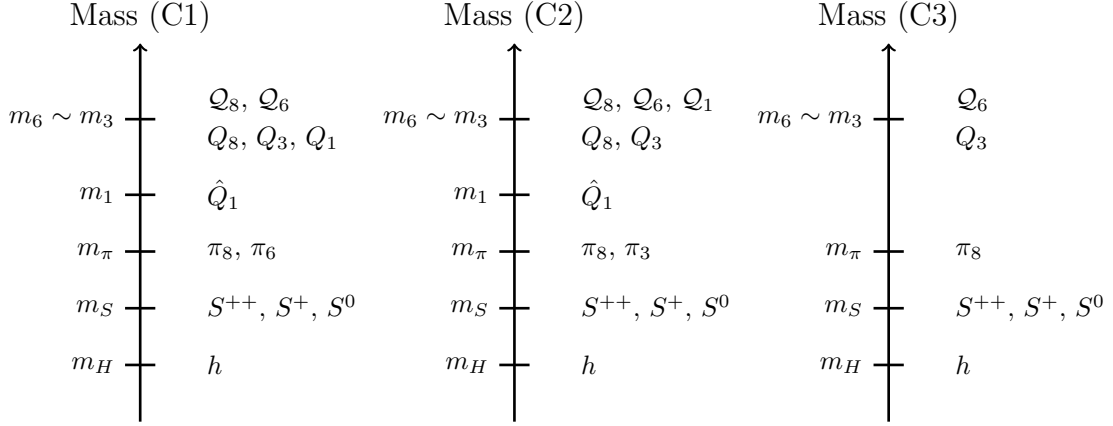
\begin{figure}
    \centering
    \begin{tikzpicture}[line width=1pt]
    
    % Vertical axis with arrow
    \draw[->] (0,1) -- (0,6) node[above] {\large Mass (C1)};
    
    % Tick marks and labels
    \foreach \y/\label in {
        5/{m_6 \sim m_3},
        4/{m_1},
        3.25/{m_\pi},
        2.5/{m_S},
        1.75/{m_H}
    } {
        \draw (-0.2,\y) -- (0.2,\y);
        \node[left] at (-0.2,\y) {$\label$};
    }
    
    % Right-side annotations
    \node[right] at (0.75,5.25) {$\mathcal Q_8,\, \mathcal Q_6$};
    \node[right] at (0.75,4.75) {$Q_8,\, Q_3,\, Q_1$};
    \node[right] at (0.75,4) {$\hat{Q}_1$};
    \node[right] at (0.75,3.25) {$\pi_8,\, \pi_6$};
    \node[right] at (0.75,2.5) {$S^{++},\, S^{+},\, S^0$};
    \node[right] at (0.75,1.75) {$h$};
    
    \end{tikzpicture}
        \begin{tikzpicture}[line width=1pt]
    
    % Vertical axis with arrow
    \draw[->] (0,1) -- (0,6) node[above] {\large Mass (C2)};
    
    % Tick marks and labels
    \foreach \y/\label in {
        5/{m_6 \sim m_3},
        4/{m_1},
        3.25/{m_\pi},
        2.5/{m_S},
        1.75/{m_H}
    } {
        \draw (-0.2,\y) -- (0.2,\y);
        \node[left] at (-0.2,\y) {$\label$};
    }
    
    % Right-side annotations
    \node[right] at (0.75,5.25) {$\mathcal Q_8,\, \mathcal Q_6,\, \mathcal Q_1$};
    \node[right] at (0.75,4.75) {$Q_8,\, Q_3$};
    \node[right] at (0.75,4) {$\hat{Q}_1$};
    \node[right] at (0.75,3.25) {$\pi_8,\,  \pi_3$};
    \node[right] at (0.75,2.5) {$S^{++},\, S^{+},\, S^0$};
    \node[right] at (0.75,1.75) {$h$};
    
    \end{tikzpicture}
            \begin{tikzpicture}[line width=1pt]
    
    % Vertical axis with arrow
    \draw[->] (0,1) -- (0,6) node[above] {\large Mass (C3)};
    
    % Tick marks and labels
    \foreach \y/\label in {
        5/{m_6 \sim m_3},
        3.25/{m_\pi},
        2.5/{m_S},
        1.75/{m_H}
    } {
        \draw (-0.2,\y) -- (0.2,\y);
        \node[left] at (-0.2,\y) {$\label$};
    }
    
    % Right-side annotations
    \node[right] at (0.75,5.25) {$\mathcal Q_6$};
    \node[right] at (0.75,4.75) {$Q_3$};
    \node[right] at (0.75,4) {};
    \node[right] at (0.75,3.25) {$\pi_8$};
    \node[right] at (0.75,2.5) {$S^{++},\, S^{+},\, S^0$};
    \node[right] at (0.75,1.75) {$h$};
    
    \end{tikzpicture}
    \caption{Schematic mass hierarchy of the pNGBs and hyperbaryons in the three classes of models C1 (left), C2 (middle) and C3 (right).}
    \label{fig:masses}
\end{figure}

The collider phenomenology of these models depends crucially on the mass hierarchy among the lightest composite states, which in the present case consist of the spin-0 pNGBs and the spin-$1/2$ hyperbaryons. The relevant hierarchies for the models of classes C1, C2, and C3 are sketched in \cref{fig:masses} (see also \cref{tab:modelclasses}). The lightest states are the electroweak pNGBs from the $\SU(5)/\SO(5)$ coset, which include the observed $125$~GeV Higgs boson $h$ together with a set of heavier states that we collectively denote by $S$. For pNGB decay constants of order $1$~TeV, their masses are naturally expected to lie in the few-hundred-GeV range. 

The QCD-coloured pNGBs, namely $\pi_8$, $\pi_6$ and $\pi_3$, are expected to be heavier because QCD radiative corrections contribute significantly to their masses, so that they typically lie around the TeV scale. The baryonic resonances are in turn expected to be heavier than the pNGBs and to populate the multi-TeV region. Each hyperbaryon multiplet comprises nearly degenerate colour representations: for models of class C1, $\mathcal{B}_S \equiv (\mathcal{Q}_8,\, \mathcal{Q}_6)$, $\mathcal{B}_A \equiv (Q_8,\, Q_3,\, Q_1)$ and $\mathcal{B}_1 \equiv \hat{Q}_1$; for models of class C2, $\mathcal{B}_S \equiv (\mathcal{Q}_8,\, \mathcal{Q}_6,\, \mathcal{Q}_1)$, $\mathcal{B}_A \equiv (Q_8,\, Q_3)$ and $\mathcal{B}_1 \equiv \hat{Q}_1$; while for models of class C3, $\mathcal{B}_3 \equiv Q_3$ and $\mathcal{B}_6 \equiv \mathcal{Q}_6$. The colour triplets mix with the top quark through partial compositeness, as shown in \cref{eq:PC_general}, so that the components carrying the same quantum numbers as the SM top and bottom quarks receive positive mass shifts of order a few hundred GeV. Additional splittings among the other mass eigenstates are induced radiatively, with QCD loops generating percent-level effects and electroweak loops contributing at the permille level~\cite{Cacciapaglia:2021uqh}. By contrast, the masses of the different hyperbaryon multiplets are set directly by the hypercolour dynamics and can therefore differ substantially. In the following, for simplicity, we assume that the multiplets containing triplets and sextets, whose masses are denoted by $m_3$ and $m_6$ in all classes of models, are approximately degenerate and heavier than the singlet multiplet of mass $m_1$ whenever the latter is present,
\begin{align}
    \text{C1 and C2}: \  m_{6} \sim m_3 > m_1\,, \qquad
    \text{C3}: \  m_6 \sim m_3\,.
\end{align}

\begin{figure}
    \centering
    \includegraphics[width=0.45\linewidth]{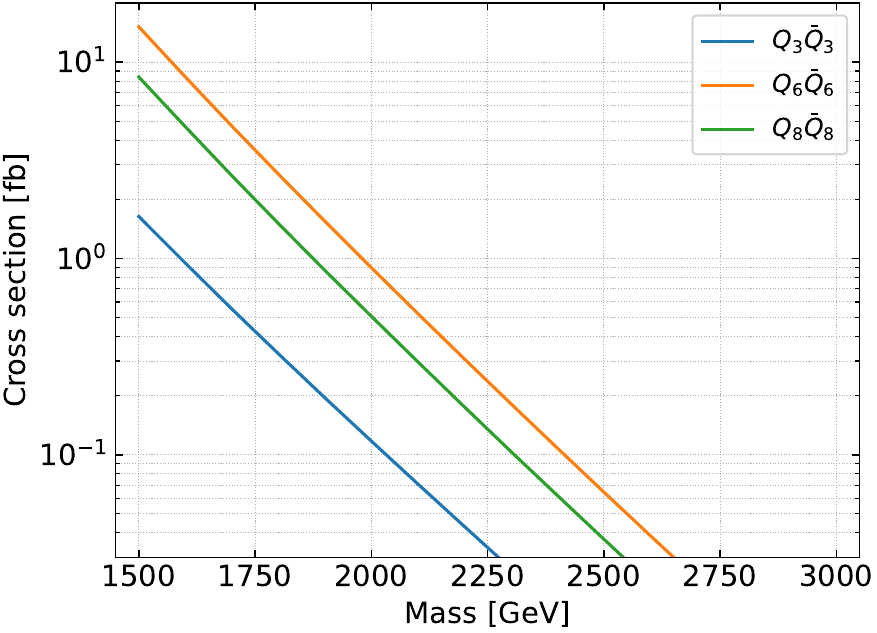}
    \includegraphics[width=0.45\linewidth]{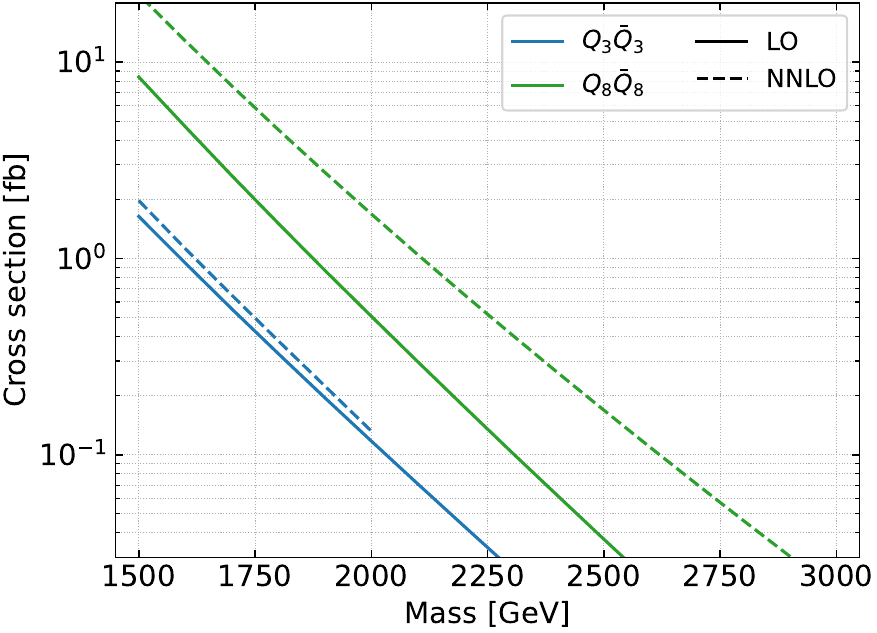}
    \caption{Pair-production cross sections for fermions in different colour representations, at leading-order for colour triplet, sextet and Majorana octet fermions (left), as well as in the context of a comparison with higher-order calculations where available (right).}
    \label{fig:crosssections}
\end{figure}

A remarkable consequence of this mass pattern is that for equal masses, the sextets have a larger pair-production cross section than the triplets because of the enhancement from the QCD colour factors. This behaviour is illustrated in \cref{fig:crosssections}, where we present hyperbaryon pair-production cross sections in proton-proton collisions at a centre-of-mass energy $\sqrt s=13$~TeV, computed with renormalisation and factorisation scales fixed to $\mu_R=\mu_F=m_i$. The leading-order cross sections are obtained with \texttt{MadGraph5\_aMC@NLO} v3.6.6~\cite{Alwall:2014hca}, using the \texttt{NNPDF2.3} leading-order set of parton densities~\cite{Ball:2012cx} as provided by \texttt{LHAPDF}~\cite{Buckley:2014ana} and a UFO~\cite{Degrande:2011ua, Darme:2023jdn} model generated with \texttt{FeynRules}~\cite{Christensen:2009jx, Alloul:2013bka}. In the left panel we compare the cross sections for the production of a pair of colour-triplet, colour-sextet and Majorana colour-octet fermions. In addition, in our models the cross section for a full hyperbaryon multiplet should be larger by a factor of five because of the multiplicity of the states within the electroweak coset. In the right panel of the figure, we compare these leading-order results to higher-order predictions available at NNLO+NNLL accuracy for a colour-octet Majorana fermion~\cite{Beenakker:2016lwe} and for a colour-triplet heavy fermion~\cite{Czakon:2011xx}. To the best of our knowledge, analogous higher-order calculations are not yet available for colour-sextet fermions.

In the subsections below, we first summarise the decays of and current bounds on the states lighter than the sextets (\cref{sec:phenolight}), which may appear in their decay chains. We do not consider decays into other coloured hyperbaryons, as these are expected to be nearly degenerate in mass so that the associated decays are suppressed or forbidden. We then turn to the characteristic decay patterns of the sextets themselves (\cref{sec:sextetdecays}), briefly comment on their single production modes (\cref{sec:sextetsingle}), and finally discuss the current LHC sensitivity together with the projected reach at its HL-LHC phase (\cref{sec:LHC}). This focus is motivated by two considerations. First, unlike the standard colour-triplet top partners, colour-sextet fermions have received comparatively little attention in the literature despite being a robust prediction of several UV-complete composite Higgs constructions. Second, for equal masses their pair-production cross sections are enhanced by QCD colour factors, making them particularly promising targets for hadron-collider searches.

\subsection{Lighter states relevant for the sextet phenomenology}\label{sec:phenolight}
In this subsection, we summarise the phenomenological properties of the states lighter than the sextets that may appear in their decay chains. These include the electroweak pNGBs from the $\SU(5)/\SO(5)$ coset, the coloured pNGBs, the light singlet fermion $\hat Q_1$, and, in some cases, the lightest colour-triplet or colour-octet hyperbaryons.

The electroweak pNGBs arise from the coset $\SU(5)/\SO(5)$~\cite{Dugan:1984hq, Ferretti:2014qta, Agugliaro:2018vsu}. Besides the observed Higgs boson, they include a doubly charged scalar $S^{++}$, two singly charged states $S_i^+$ and four additional neutral scalars which however do not contribute to the processes that we study here. For simplicity, we assume these states to be approximately mass degenerate, and that their couplings to third-generation quarks dominate. This implies that they mostly decay through\footnote{If the couplings to quarks are absent, the dominant decays involve pairs of electroweak gauge bosons ($W^\pm$, $Z$ and $\gamma$), leading to bounds in the range $400-700$~GeV~\cite{Cacciapaglia:2022bax}.}
\begin{align}
    S^+ \to t \bar b
    \qquad\text{and}\qquad 
    S^{++} \to W^+ S^{+(*)} \to W^+ t \bar b\,.
\end{align}
Since these states are pair-produced dominantly through electroweak interactions~\cite{Agugliaro:2018vsu}, current non-dedicated LHC searches exclude masses up to about $m_S \sim 450$~GeV~\cite{Cacciapaglia:2022bax}. In the following, we therefore fix
\begin{align}
    m_S = 500~\text{GeV}\,.
\end{align}
The coloured pNGBs are also assumed to decay predominantly to third-generation quarks, leading to the main decay modes\footnote{For the triplet state $\pi_3$, the strange quark must be involved in the decay because the $\pi_3$ couplings to quarks are antisymmetric in the flavour space. In addition, the couplings of $\pi_6$ and $\pi_3$ violate the baryon number, and the triplet can also decay into a top quark and a lepton, thus violating lepton number.}
\begin{align}
    \pi_8 \to t\bar t\,, \qquad
    \pi_6 \to bb
    \qquad\text{and}\qquad 
    \pi_3 \to \bar b \bar s\,.
\end{align}
Among these channels, the strongest bound typically comes from the ubiquitous pair production of the octet, implying $m_\pi \gtrsim 1.375~\text{TeV}$ from current searches~\cite{Kunkel:2025qld}. In what follows, we again assume the coloured pNGBs to be approximately degenerate and then choose the benchmark value
\begin{align}\label{eq:octetpNGBbound}
    m_\pi = 1.5~\text{TeV}\,.
\end{align}

The light singlet multiplet $\hat Q_1$ contains one charged state $\hat Q_1^\pm$, one neutral Dirac state $\hat Q_1^0$ and one neutral Majorana state $\hat q_1^0$. Electroweak interactions induce small mass splittings at the permille level, such that the lightest state is the Majorana fermion $\hat q_1^0$ which could also serve as a dark matter candidate. The heavier members of the multiplet therefore decay to it via off-shell $W$ and $Z$ bosons, schematically as
\begin{align}\label{eq:decayQ10}
    \hat Q_1^{+,0} \to \hat q_1^0 + \text{soft objects}\,.
\end{align}
Since the visible decay products are too soft to be detected efficiently at the LHC, all $\hat Q_1$ states may in practice be treated as detector-stable. Moreover, since pair production proceeds only through electroweak interactions, the resulting constraints are at best weak or even absent. Among the colour-triplet top partners, the state $X_{5/3}$ plays a particularly important role for the sextet phenomenology. It is typically the lightest state in the triplet multiplet and, due to its exotic electric charge, it does not mix directly with the SM quarks~\cite{Bizot:2018tds, Xie:2019gya}. Partial compositeness instead induces decays into electroweak pNGBs,
\begin{align}\label{eq:x53toscalars}
    X_{5/3} \to b S^{++}\qquad\text{or}\qquad X_{5/3} \to t S^+\,,
\end{align}
with fixed relative branching ratios~\cite{Cacciapaglia:2021uqh} in the ratio \mbox{$2\!:\!\!1$}. In addition, the standard decay into SM particles,
\begin{align}\label{eq:x53towt}
    X_{5/3} \to W^+ t\,,
\end{align}
is allowed through the $X_{5/3}$-$W$-$X_{2/3}$ kinetic coupling combined with the mixing of the $X_{2/3}$ state with the top quark. However, this channel is suppressed by Higgs insertions, so that its partial width scales as $(v/f_\psi)^4 \lesssim 10^{-3}$. No such suppression affects the scalar decays of \cref{eq:x53toscalars}, which therefore typically dominate. An exception may occur in strongly compressed spectra, where the lighter final state system of \cref{eq:x53towt} becomes kinematically favoured. Finally, in models C1 and C2, both the $Q_8$ and $\mathcal Q_8$ colour-octet hyperbaryons can decay into the singlet $\hat Q_1$ together with a $\pi_8$ pNGB through derivative couplings, leading to a representative lower bound around $2.7$~TeV~\cite{Cacciapaglia:2021uqh}. In class C2, an additional decay mode of the $Q_8$ hyperbaryon into top or bottom quarks plus the triplet pNGB $\pi_3$ is also allowed, yielding similar constraints. In constructions of classes C1 and C2, the sextet masses are naturally tied to those of the $\mathcal Q_8$ states, whereas this connection is absent in C3. For this reason, in the following we will also consider sextet masses below $2.7$~TeV.

\subsection{Decay patterns of the sextet hyperbaryons}\label{sec:sextetdecays}
As discussed in \cref{sec:sextets}, in all models the sextets couple to the triplets through the colour-octet pNGB, as shown in \cref{eq:Lag368}. This leads schematically to decay processes of the form
\begin{align}
    \mathcal Q_6 \to Q_3^c\,\pi_8\,.
\end{align}
However, if the sextet and triplet masses are comparable, these decays are often kinematically closed, since the octet pNGB is already constrained to be rather heavy, with $m_{\pi_8} \gtrsim 1.375$~TeV~\cite{Kunkel:2025qld}. In practice, the relevant decay modes therefore proceed either through the mixing of the triplets with the SM quarks or, when present, through direct couplings induced by partial compositeness. Taking into account the electroweak quantum numbers, the generic sextet decay pattern is therefore expected to be 
\begin{align} \label{eq:Q6pi8Q3_decays}
    \mathcal Q_6^{-5/3} \to (\bar X_{5/3} \pi_8)^\ast\,, \qquad 
    \mathcal Q_6^{-2/3} \to \bar t \pi_8
    \qquad\text{and}\qquad 
    \mathcal Q_6^{1/3} \to \bar b \pi_8\,.
\end{align}
A direct two-body decay is absent for the $\mathcal Q_6^{-5/3}$ state because of electric-charge conservation, so this state necessarily decays through an off-shell $X_{5/3}$ or $\pi_8$ state. Before discussing this case in more detail, let us stress that not all partial widths of the states with charge $-2/3$ and $1/3$ are expected to be comparable. In particular, there are three distinct $\mathcal Q_6^{-2/3}$ states, which we label according to the corresponding triplets, as introduced in \cref{eq:nomenclature}. The state $\mathcal Q_{6,X}^{-2/3}$ couples to the top only through partial-compositeness mixings, yielding a suppression factor $v^2/f^2$. As a result, the partial width for the decay into a $\bar t\,\pi_8$ system is suppressed by a factor $v^4/f^4\lesssim 10^{-3}$ with respect to the widths of the other $\mathcal Q_{6,TL}^{-2/3}$, $\mathcal Q_{6,TR}^{-2/3}$ and $\mathcal Q_6^{1/3}$ states. This suppression is phenomenologically important, since it allows the $\mathcal Q_{6,X}^{-2/3}$ three-body decays to compete, while the corresponding two-body modes remain dominant for the other sextets. Finally, additional contributions to the decays listed in \cref{eq:Q6pi8Q3_decays} may arise from direct couplings to top quarks induced by partial compositeness (see \cref{eq:PC_sextet}), whenever such interactions are present.

A special case is the state $\mathcal Q_6^{-5/3}$. Its two-body decay $\mathcal Q_6^{-5/3}\to \bar X_{5/3}\pi_8$ is kinematically open only for sextets much heavier than the triplets, namely for $m_6\gtrsim 3$~TeV once the current lower limits on $X_{5/3}$ and $\pi_8$ discussed in \cref{sec:phenolight} are taken into account. Since this region may already be difficult to probe due to low production rates (as shown in \cref{fig:crosssections}), we focus instead on the $\mathcal Q_6^{-5/3}$ three-body decays mediated by an off-shell $X_{5/3}$ or $\pi_8$ state. In this case, the coupling controlling the decay $X_{5/3}\to tW^+$ is suppressed by $v^2/f^2$, whereas those relevant for the decays $X_{5/3}\to bS^{++}$ and $X_{5/3}\to tS^+$ are not. In addition, the coupling appearing in the $\pi_8\to t\bar t$ partial width is itself suppressed by $v/f$~\cite{Cacciapaglia:2015eqa}. For sufficiently heavy sextets, the dominant channels are therefore expected to be
\begin{equation}\label{eq:q653_decays}\begin{split}
  \mathcal Q_6^{-5/3}\to \bar t\,S^-\,\pi_8 \qquad &\text{for } m_6\gtrsim 2~\text{TeV}\,, \\
  \mathcal Q_6^{-5/3}\to \bar b\,S^{--}\,\pi_8 \qquad &\text{for } m_6\gtrsim 1.8~\text{TeV}\,,
\end{split}\end{equation}
once the lower limits on the masses of the $S$ and $\pi_8$ pNGBs are taken into account. For lighter sextet masses or for heavier electroweak pNGBs, only the most suppressed channel remains open,
\begin{align}\label{eq:q653_decay_low_mass}
    \mathcal Q_6^{-5/3}\to \bar t\,W^-\,\pi_8 \qquad \text{for } m_6\gtrsim 1.6~\text{TeV}\,.
\end{align}
Finally, the channel $\mathcal Q_6^{-5/3}\to \bar X_{5/3}\,\bar t\,t$ mediated by an off-shell $\pi_8$ pNGB could also become relevant for a sufficiently light $X_{5/3}$, as may occur in models of class C3. Even in that case, however, the mass limit on $X_{5/3}$ typically requires $m_6\gtrsim 1.85$~TeV, so that this mode remains subleading compared with the decay channels through the electroweak pNGBs.

We now turn to the model-dependent modifications of the generic pattern introduced above. In models of class C1, additional two-body decays are allowed through the sextet pNGBs $\pi_6$ and the singlet hyperbaryons $\hat Q_1$,
\begin{align}\label{eq:decay_q6pi6q1}
    \mathcal Q_6 \to \hat Q_1\, \pi_6\,,
\end{align}
provided the $\hat Q_1$ states are sufficiently light. These decays are expected to dominate for the $\mathcal Q_6^{-5/3}$ and $\mathcal Q_{6,X}^{-2/3}$ states, whose alternative channels are suppressed either by phase space or by small couplings. One then has explicitly
\begin{equation}\begin{split}
    \mathcal Q_6^{-5/3}\to \hat Q_1^{-1}\,\pi_6\qquad\text{and} \qquad
    \mathcal Q_{6,X}^{-2/3}\to \hat Q_1^0\,\pi_6\,.
\end{split}\end{equation}
For the other sextets, decays into $\hat Q_1$ can compete with the generic channels of \cref{eq:Q6pi8Q3_decays},
\begin{align}
    \mathcal Q_{6,TL}^{-2/3}\to \hat Q_1^0\,\pi_6\,, \qquad
    \mathcal Q_{6,TR}^{-2/3}\to \hat q_1^0\,\pi_6\qquad\text{and} \qquad
    \mathcal Q_6^{1/3}\to \hat Q_1^{1}\,\pi_6\,.
\end{align}
After the subsequent decay $\pi_6\to bb$, and since all states in the $\hat Q_1$ multiplet contribute only to missing transverse energy $E_T^\mathrm{miss}$ (including soft undetected particles), as shown in \cref{eq:decayQ10}, these channels lead to the common detector-level signature
\begin{align}
    \mathcal Q_6 \to bb + E_T^\mathrm{miss}\,.
\end{align}
Finally, models of classes C1 and C2 also allow decays into the colour octets $Q_8$ hyperbaryon, mediated by intermediate $\pi_6$ and $\pi_3$ pNGBs,
\begin{equation}
    \mathcal Q_6 \to \pi_6\,Q_8^\ast \to \pi_6\,\pi_8\,\hat Q_1 \quad \text{for C1}
    \qquad\text{and\qquad}
    \mathcal Q_6 \to \pi_3^c\,Q_8^\ast \to \pi_3^c\,\pi_8\,\hat Q_1 \quad \text{for C2}\,.
\end{equation}
Since the octets are constrained to be too heavy to be produced on shell, these channels can only proceed as three-body decays through an off-shell $Q_8$ fermion. In view of the lower limits on the coloured pNGB masses, they nevertheless become kinematically accessible only for rather heavy sextets, with masses roughly above $3$~TeV, and we therefore do not consider them further.

\begin{table}\renewcommand{\arraystretch}{1.4}
\centering
\begin{tabular}{c|cc|cc}
 & \multicolumn{2}{c|}{All models} & \multicolumn{2}{c}{C1}  \\ \hline
$\mathcal{Q}_{6,B}^{1/3}$ & $\bar{b} \pi_8 \to \bar{b} \bar{t} t$ & dominant & $\hat{Q}_1^{1} \pi_6 \to bb+\mbox{MET}$ & competitive  \\[.2cm]
$\mathcal{Q}_{6,TL}^{-2/3}$ & $\bar{t} \pi_8 \to \bar{t} \bar{t} t$ & dominant & $\hat{Q}_1^0 \pi_6 \to bb+\mbox{MET}$ & competitive  \\[.2cm]
$\mathcal{Q}_{6,TR}^{-2/3}$ & $\bar{t} \pi_8 \to \bar{t} \bar{t} t$ & dominant & $\hat{q}_1^0 \pi_6 \to bb+\mbox{MET}$ & competitive  \\[.2cm] 
$\mathcal{Q}_{6,X}^{-2/3}$ & $\bar{t} \pi_8 \to \bar{t} \bar{t} t$ & suppressed & $\hat{Q}_1^{0} \pi_6 \to bb+\mbox{MET}$ & dominant  \\[.2cm] 
\multirow{3}{*}{$\mathcal{Q}_{6,X}^{-5/3}$} & $\bar{t} S^- \pi_8 \to \bar{t} \bar{t} b\bar{t} t$ & 3-body & \multirow{3}{*}{$\hat{Q}_{1}^{-1} \pi_6 \to bb+\mbox{MET}$} & \multirow{3}{*}{dominant}  \\
 & $\bar{b} S^{--} \pi_8 \to \bar{b} \bar{t} b W^-\bar{t} t$ & 3-body &  & \\ 
  & $\bar{t} W^{-} \pi_8 \to \bar{t} W^-\bar{t} t$ & 3-body, heavy $S$ &  & \\
\end{tabular}
\caption{\label{tab:decaysQ6} Summary of the relevant decay modes of the sextet hyperbaryons $\mathcal Q_6$ for masses below $3$~TeV. We indicate the generic channels present in all model classes, together with the additional modes specific to class C1, and indicate whether they are expected to be dominant, competitive or suppressed under the benchmark assumptions described in the text.}
\end{table}

The decay patterns of the sextets $\mathcal Q_6$ for masses below $3$~TeV are summarised in \cref{tab:decaysQ6}.

\subsection{Sextet single production prospects}\label{sec:sextetsingle}
Besides QCD pair production with cross sections shown in \cref{fig:crosssections} and depending essentially only on the sextet mass, the states of charge $-2/3$ and $1/3$ may also be produced singly through their couplings to top and bottom quarks. 

A first class of processes proceeds through the ubiquitous colour-octet pNGB $\pi_8$ and the interaction of \cref{eq:Lag368}. Since the $\pi_8$ state can itself be produced from gluon fusion through its effective coupling to two gluons induced by top loops and the hypercolour anomaly~\cite{Belyaev:2016ftv}, one may have associated the sextet single-production channels
\begin{align}
    gg \to \pi_8^\ast \to \mathcal{Q}_6^{-2/3}\, t \qquad\text{and}\qquad
    gg \to \pi_8^\ast \to \mathcal{Q}_6^{1/3}\, b\,.
\end{align}
These channels are, however, expected to be suppressed both by the loop-induced $\pi_8$-$g$-$g$ coupling and, in the kinematic regime relevant for our benchmarks, by the off-shellness of the $\pi_8$ propagator. A second class of processes can instead be described in terms of an effective chromomagnetic interaction with gluons. Such a coupling is generated at one loop, with a $\pi_8$ running in the loop and an external gluon emitted, and is analogous to the chromomagnetic single production mechanisms discussed for colour-triplet top partners in the literature~\cite{Belyaev:2021zgq, Belyaev:2022ylq, Belyaev:2023yym}. At the effective level, the relevant interactions may be parametrised as
\begin{align}
    \frac{g_s \kappa_t}{16\pi^2 f_\chi}\,\mathcal{Q}_6^{-2/3}\sigma^{\mu\nu} t\, G_{\mu\nu}
    + \frac{g_s \kappa_b}{16\pi^2 f_\chi}\,\mathcal{Q}_6^{1/3}\sigma^{\mu\nu} b\, G_{\mu\nu}\,,
\end{align}
where $g_s$ is the QCD gauge coupling and $\kappa_{t,b}$ denote loop functions that also encode the mixing of the triplet top partners with the top and bottom fields. These operators induce the associated production of sextets from both gluon-pair and quark-antiquark initial states,
\begin{align}
    q\bar q,\; gg \to \mathcal{Q}_6^{-2/3}\, t
    \qquad\text{and}\qquad
    q\bar q,\; gg \to \mathcal{Q}_6^{1/3}\, b\,.
\end{align}
Since these contributions are not additionally suppressed by an off-shell $\pi_8$ propagator, they may become comparable to, or larger than, the previous class of processes, depending on the size of the effective coefficients.

Compared with QCD pair production, single production is suppressed by model-dependent couplings and loop factors, but benefits from a milder phase-space suppression, especially in light of current bounds. It is therefore expected to become relevant mainly for heavy sextet masses, close to the kinematic limit of pair production. For this reason and because these channels are expected to become most relevant in the HL-LHC regime, we leave a dedicated analysis of single sextet production to future work.

\section{LHC bounds and HL-LHC projections}
\label{sec:LHC}

So far, to the best of our knowledge, no dedicated LHC search has targeted pair-produced colour-sextet fermions. Nevertheless, this mode often leads to final states with large jet multiplicities, multiple $b$-jets and in some cases sizeable missing transverse energy. In the meantime, existing searches designed for supersymmetry are therefore expected to be useful, and can be recast to constrain the scenarios considered here and assess the sensitivity of the LHC to them. 

We generate hard-scattering events using the same setup as for \cref{fig:crosssections}. Since no next-to-leading-order calculation is currently available for sextet pair production, we present results both at leading order and after applying a representative $K$-factor of $K=3$. This choice is motivated by the sizeable higher-order corrections observed for colour-octet fermion production, and we use it here as an estimate of the potential theory uncertainty. The parton-level events are next passed to \texttt{Pythia8}~\cite{Sjostrand:2014zea} for parton showering and hadronisation, producing output in the \texttt{HepMC} format~\cite{Dobbs:2001ck}. A technical complication arises because current public showering tools do not straightforwardly support the colour flow associated with a $\mathbf3 \otimes \mathbf6 \otimes \mathbf8$ vertex in the colour space. For processes involving such couplings, we therefore adopt a surrogate model, replacing the sextet fermion by a triplet fermion using the public \texttt{eVLQ} implementation~\cite{Banerjee:2022izw} in \texttt{FeynRules}, and then rescaling the total production rate to the sextet pair-production one. This approximation cannot be validated exactly within the present tool chain. However, we expect it to capture the dominant effects relevant for the searches considered here since the most sensitive analyses rely primarily on inclusive observables such as jet and $b$-jet multiplicities and missing transverse energy, rather than on jet-substructure observables that would be more directly sensitive to the detailed colour flow. The showered events are subsequently analysed with \texttt{MadAnalysis5}\footnote{We use the \texttt{substructure} branch: \url{https://github.com/MadAnalysis/madanalysis5/tree/substructure}, commit \texttt{75fbc8f}.}~\cite{Conte:2012fm, Conte:2014zja, Dumont:2014tja, Conte:2018vmg, Araz:2025bww}. Jets are clustered with the anti-$k_T$ algorithm~\cite{Cacciari:2008gp} as implemented in \texttt{FastJet}~\cite{Cacciari:2011ma}, while detector effects are modelled either with \texttt{Delphes 3}~\cite{deFavereau:2013fsa} or with the \texttt{MadAnalysis5} simplified fast simulation (SFS) framework~\cite{Araz:2020lnp, Araz:2021akd}. We then pass the events through all recast analyses available in the public analysis database of \texttt{MadAnalysis}.\footnote{See \url{https://madanalysis.irmp.ucl.ac.be/wiki/PublicAnalysisDatabase}.} For each signal region of each analysis, the number of surviving events is used to derive exclusion limits via the CL$_s$ method~\cite{Read:2002hq} as embedded in \spey~\cite{Araz:2023bwx} (that also handles simplified likelihoods and \pyhf~\cite{Heinrich:2021gyp} statistical models when available), and we use the expected limit to identify the most sensitive signal region in each search and report the corresponding observed limit as our final result. 

\begin{figure}[t]
    \centering
    \includegraphics[width=0.45\linewidth]{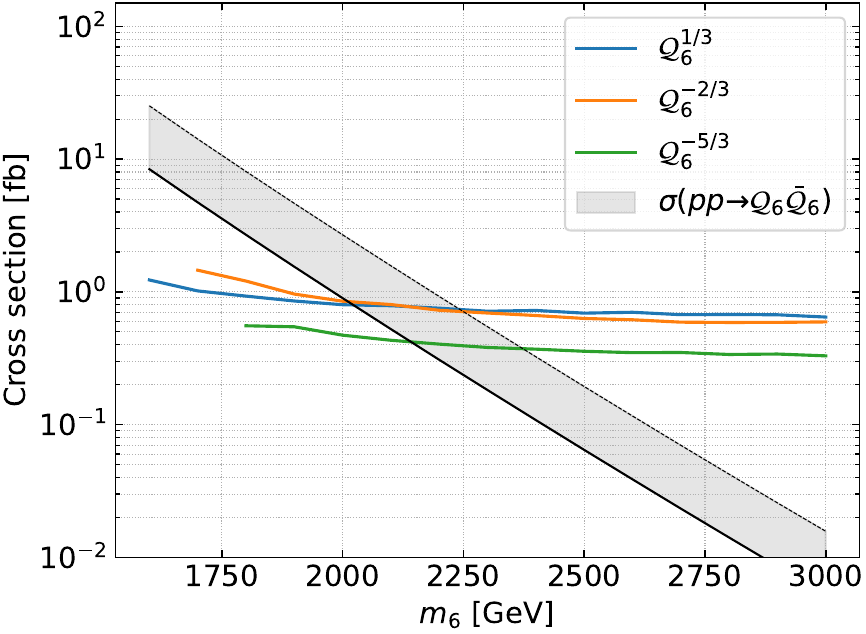}\hfill
    \includegraphics[width=0.45\linewidth]{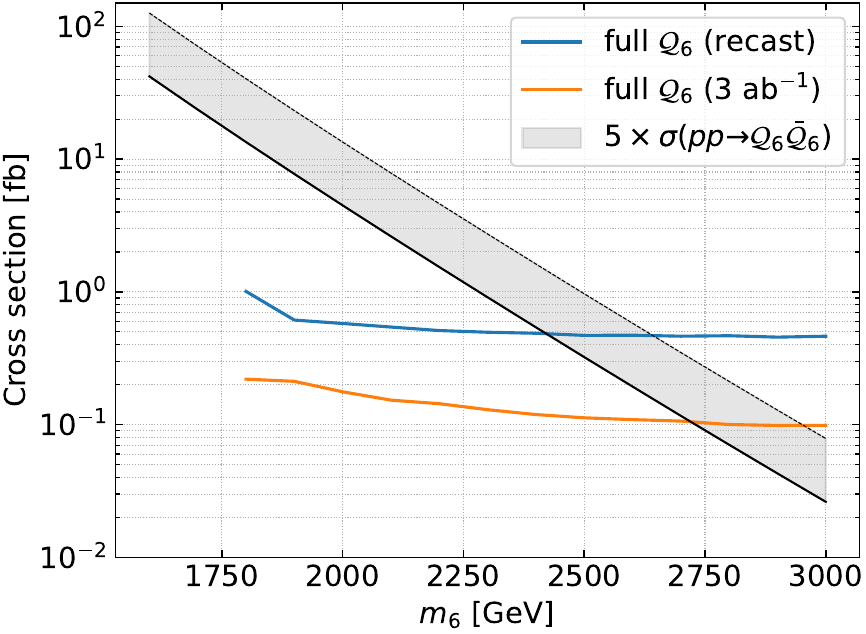}
    \caption{Recast bounds on the sextet hyperbaryons $\mathcal Q_6$ decaying through the octet pNGB $\pi_8$. The grey band shows the QCD pair-production cross section, with the lower edge corresponding to the leading-order prediction and the upper edge to the same prediction rescaled by a representative $K$-factor of $3$. In the left panel, the limits are shown separately for the three relevant sextet charge states, and compared to the single-component pair-production cross section $\sigma(pp \to \mathcal{Q}_6 \bar{\mathcal{Q}}_6)$. In the right panel, the corresponding bound is shown for the full sextet multiplet, and compared to five times the single-component cross section, $5\times\sigma(pp \to \mathcal{Q}_6 \bar{\mathcal{Q}}_6)$, reflecting the five charge eigenstates contributing to the total production rate. The blue curve shows the bounds obtained from the current LHC dataset, while the orange curves shows the extrapolated HL-LHC sensitivity at $3~\mathrm{ab}^{-1}$ and $\sqrt{s}=13$~TeV.}
    \label{fig:368bounds}
\end{figure}

Using the workflow described above, we derive 95\% CL exclusion limits on the pair-production cross section of the sextets. We present the leading-order prediction together with a shaded band obtained by rescaling the rate by a representative factor of $K=3$. This band should not, however, be interpreted as a precise estimate of the higher-order correction, but rather as an indication of the potentially sizeable theory uncertainty associated with the leading-order prediction. In the left panel of \cref{fig:368bounds} we show the bounds obtained for the generically expected decay chains
\begin{align}
    \mathcal Q_6^{1/3} \to \bar b \pi_8 \to \bar b \bar t t, \qquad
    \mathcal Q_6^{-2/3} \to \bar t \pi_8 \to \bar t \bar t t, \qquad
    \mathcal Q_6^{-5/3} \to \bar t W^- \pi_8 \to \bar t W^- \bar t t,
\end{align}
which were derived in \cref{sec:pheno} and collected in the left column of \cref{tab:decaysQ6}. We recall that throughout this study, the octet pNGB mass is fixed to $m_{\pi_8}=1.5~\mathrm{TeV}$, as depicted in \cref{eq:octetpNGBbound} and in light of current limits. The dominant exclusions are obtained from the recasts of the \texttt{ATLAS-SUSY-2018-31}~\cite{ATLAS:2019gdh, IHALED_2020} and \texttt{ATLAS-SUSY-2018-17}~\cite{ATLAS:2020xgt, I2CZWU_2021} searches. These are particularly relevant because they target final states with high jet multiplicities, multiple $b$-jets and a sizeable amount of missing transverse momentum, closely resembling the signatures generated by sextet pair production. For the states with an electric charge of $-5/3$ we include only the $\mathcal Q_6^{-5/3}\to \bar t W^- \pi_8$ decay channel since this is the only one expected to be open over the full low-mass range of interest. At larger sextet masses, the additional modes $\mathcal Q_6^{-5/3}\to \bar t S^- \pi_8$ and $\mathcal Q_6^{-5/3}\to \bar b S^{--} \pi_8$ become kinematically accessible, and their subsequent decays lead to even larger hadronic multiplicities. They are therefore expected to populate efficiently the signal regions of the searches used in our recast framework, so that the bounds obtained from the $t W \pi_8$ channel alone should be regarded as conservative.

Without (with) the representative $K$-factor of $K=3$, the states with electric charges $+1/3$ and $-2/3$ are excluded up to $2.02$ ($2.25$)~TeV, while the $\mathcal{Q}_6^{-5/3}$ state is excluded up to $2.14$ ($2.37$)~TeV. The slightly stronger bound on $\mathcal{Q}_6^{-5/3}$ relative to the other charge states is consistent with the larger visible multiplicity of its decay chain due to the additional $W$ boson, $\mathcal{Q}_6^{-5/3} \to \bar{t}\,W^-\,\pi_8$. In the fully hadronic mode, this contributes with four extra jets to the final state, thereby enhancing the efficiency of searches targeting high jet and $b$-jet multiplicities. All individual exclusion curves display a characteristic plateau at high sextet masses, where the excluded cross section becomes approximately constant. This indicates that, once the sextet decay products are sufficiently energetic, the acceptance of the most sensitive inclusive signal regions varies only mildly with $m_6$. In this regime, the mass reach is therefore driven mainly by the rapidly falling QCD pair-production cross section rather than by a strong degradation of the sensitivity. As a consequence, the intersection between the exclusion curves and the predicted production rate depends noticeably on the assumed normalisation of the sextet pair-production cross section, motivating the comparison between the leading-order result and the representative $K=3$ rescaling.

In the right panel of \cref{fig:368bounds}, we consider the full sextet multiplet simultaneously, namely one $\mathcal{Q}_6^{+1/3}$, three $\mathcal{Q}_6^{-2/3}$ and one $\mathcal{Q}_6^{-5/3}$ states all produced and decaying concurrently. The total pair-production cross section is accordingly enhanced by a factor of five relative to the single-component rate, and it is this combined rate (shown as the grey band in the figure) that should be compared to the excluded cross section for the full multiplet. Combining all components, the current LHC dataset excludes sextet masses up to $2.42$ ($2.64$)~TeV at leading order (with $K=3$), a bound that is substantially stronger than for any individual component taken in isolation. While this exclusion is comparable to the current bounds on colour-octet fermionic resonances~\cite{Cacciapaglia:2021uqh}, it should be interpreted with care. The sextet has a larger quadratic Casimir than the octet, $C_2(\mathbf 6)=10/3$ compared with $C_2(\mathbf 8)=3$, and in our setup a single Dirac sextet component has then a production rate larger than that of the Majorana octet shown in \cref{fig:crosssections}. The resulting mass reach therefore reflects the interplay between colour factors, spin and multiplicity factors, the five contributing sextet components and the different decay topologies entering the search strategies relevant in each case. On the other hand, at first sight the excluded cross section for the full multiplet may appear weaker than the strongest individual bound shown in the left panel of \cref{fig:368bounds}. This is in fact not contradictory. The individual bounds correspond to signal hypotheses in which a single sextet state decays exclusively through one topology, so that the most sensitive signal region can be closely matched to that specific final state, yielding a lower excluded cross section. The full-multiplet result, by contrast, combines several charge components with different branching patterns and therefore different selection efficiencies across signal regions, which generically results in a higher excluded cross section per component.

Finally, we extrapolate the bounds to the full HL-LHC luminosity of $3~\mathrm{ab}^{-1}$, retaining the centre-of-mass energy of $\sqrt{s}=13$~TeV as a conservative choice. While running at the design energy of $\sqrt{s}=14$~TeV is expected to slightly further improve the reach, we do not include this effect here for simplicity. We refer to Appendix~\ref{app:HLLHC} for details on the extrapolation procedure. As shown by the orange curve in the right panel of \cref{fig:368bounds}, the projected sensitivity to the full sextet multiplet reaches approximately 2.75~TeV at leading order and approaches $3$~TeV once the representative $K$-factor of $3$ is applied. These results indicate that colour-sextet fermionic resonances provide a well-motivated target for Run~3 and HL-LHC searches, and that existing analyses already have sensitivity to a substantial fraction of the multi-TeV mass range relevant for the benchmark spectra predicted by modern composite BSM models.

\begin{figure}[t]
    \centering
    \includegraphics[width=0.45\linewidth]{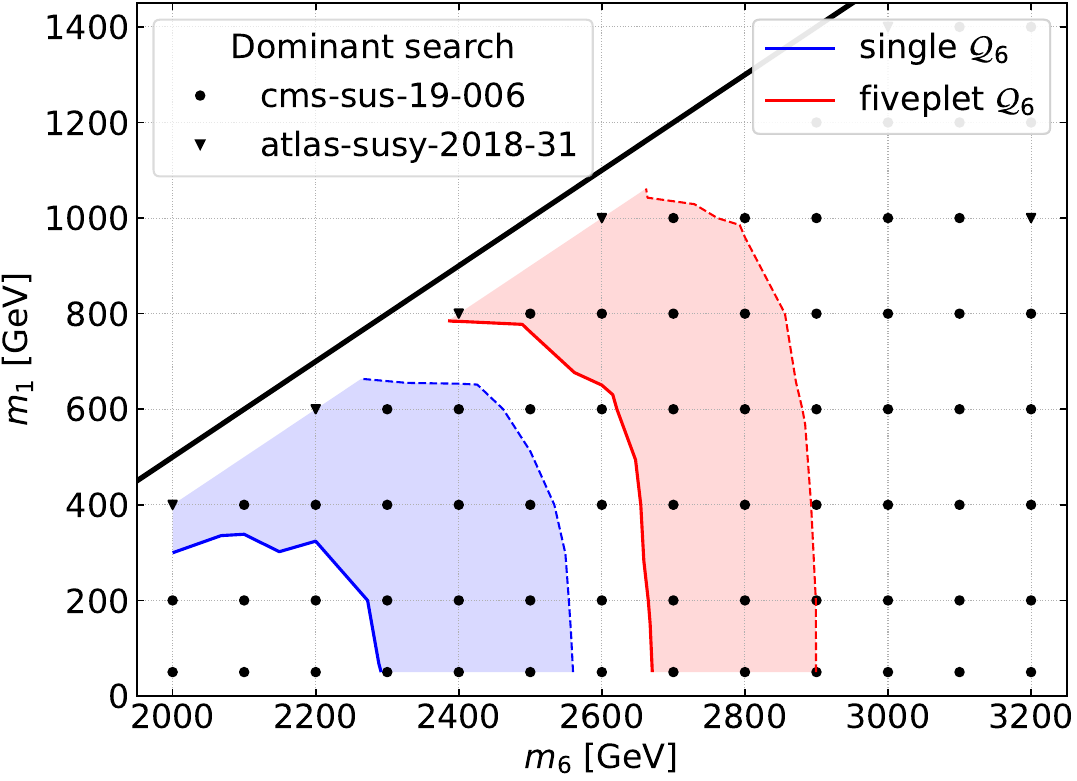}
    \includegraphics[width=0.45\linewidth]{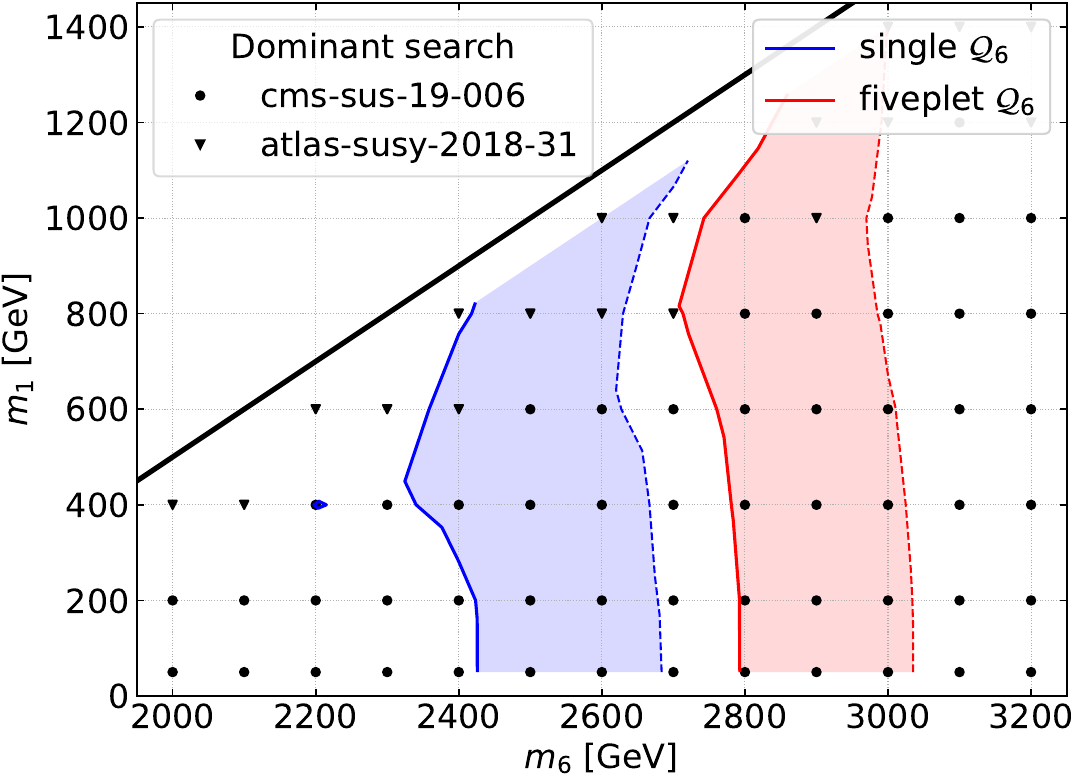}
    \caption{Recast bounds on pair-produced sextet hyperbaryons decaying through the class-C1 channel $\mathcal Q_6\to \pi_6\hat Q_1\to bb+E_T^{\rm miss}$, assuming a $100\%$ branching ratio for all five sextet components. The exclusions are shown in the $(m_6,m_1)$ plane, both for a single $\mathcal Q_6$ state (blue) and for the full multiplet (red), and the solid contours correspond to leading-order sextet pair production while the dashed ones include a $K$-factor of $3$. The left panel shows the current LHC bounds, and the right panel shows the extrapolated sensitivity to $3~\mathrm{ab}^{-1}$ at $\sqrt{s}=13$~TeV. The dominant search driving the exclusion is indicated by the markers.}
    \label{fig:q6_DM_exclusion}
\end{figure}

We also analyse the bounds arising from the decays of the $\mathcal{Q}_6$ states via the colour-sextet pNGB $\pi_6$, a channel specific to models of class C1 which is absent in models of classes C2 and C3. In this scenario, the decay chain proceeds as $\mathcal{Q}_6 \to \pi_6\,\hat{Q}_1 \to bb\,\hat{Q}_1$ where the colour-singlet baryon $\hat{Q}_1$ escapes the detector as missing transverse energy, thus yielding a $bb + E_T^{\rm miss}$ signature. This topology is especially relevant for the $\mathcal Q_6^{-5/3}$ and $\mathcal Q_{6,X}^{-2/3}$ states, for which the generic $\pi_8$-mediated channels are suppressed either by phase space or by small mixings, but it can also compete for the remaining sextet components. In our scan of the parameter space, we assume a $100\%$ branching ratio into this final state for all five sextet components, which should be understood as a benchmark choice designed to assess the maximal sensitivity to this missing-energy topology and illustrate the setup listed in the right column of \cref{tab:decaysQ6}. Although in generic strongly coupled spectra the singlet hyperbaryon $\hat Q_1$ is not expected to be lighter than the coloured pNGBs (see \cref{fig:masses}), its mass is model-dependent. We therefore keep $m_1$ as a free parameter in our scan and derive bounds in the $(m_6,m_1)$ plane, fixing the coloured pNGB mass to the benchmark value used above, $m_{\pi_6}=1.5~\mathrm{TeV}$. Below the diagonal line marking the two-body threshold, $m_6 = m_1 + m_{\pi_6}$, the decay $\mathcal Q_6\to\pi_6\hat Q_1$ is open. Above it, the same $bb+E_T^{\rm miss}$ topology can still arise through an off-shell intermediate $\pi_6$ pNGB. In this region, however, the $\pi_8$-mediated decays discussed previously are expected to dominate for most components of the multiplet, and we therefore do not consider this region as a realistic description of the dominant sextet phenomenology. 

The resulting exclusions are shown in \cref{fig:q6_DM_exclusion} (see Appendix~\ref{app:HLLHC} for details). The left panel gives the current LHC bounds, while the right panel shows the extrapolated sensitivity to $3~\mathrm{ab}^{-1}$ at fixed $\sqrt{s}=13$~TeV. The blue contours correspond to a single $\mathcal Q_6$ state, whereas the red contours include the full five-component sextet multiplet. The dominant constraints come mainly from the \texttt{CMS-SUS-19-006} search~\cite{CMS:2019zmd, 4DEJQM_2020} which targets final states with multiple jets, large $H_T$ and sizeable missing transverse momentum, and from the \texttt{ATLAS-SUSY-2018-31} analysis~\cite{ATLAS:2019gdh, IHALED_2020} which is particularly sensitive to final states with several $b$-tagged jets and missing transverse momentum. The two analyses are complementary across the mass plane. Away from the threshold, where the visible jets and missing momentum are typically harder, the inclusive jets+\(E_T^{\rm miss}\) search provides strong sensitivity. Close to the two-body threshold, the kinematics become more compressed and the multi-$b$ structure of the final state becomes comparatively more important, making the ATLAS analysis competitive or dominant in that region. As in the $\pi_8$-mediated case, considering the full multiplet substantially strengthens the bound because of the larger total production rate. Here, the current data exclude sextet masses up to about $2.65$~TeV at leading order and up to about $2.9$~TeV when the representative $K=3$ rescaling is applied. The HL-LHC extrapolation extends the reach beyond $2.8$~TeV at leading order and up to about $3.05$~TeV with the same $K$-factor. These bounds are comparable to, and in some regions stronger than, those obtained for the generic $\pi_8$-mediated decay chains. This shows that the missing-energy channel in class C1 provides an important complementary probe of sextet hyperbaryons for spectra where the singlet $\hat Q_1$ is light enough for the two-body decay to be open.

\section{Conclusions and outlook}\label{sec:outlook}

We have presented the first systematic study of the colour-sextet fermions arising in composite Higgs models with partial compositeness, focusing on the relevant model classes C1, C2 and C3 introduced in~\cref{tab:modelclasses} and classified according to the symmetry-breaking pattern of the coloured hyperquark sector. Starting from the underlying gauge theory, we have constructed the relevant interaction Lagrangians within the CCWZ framework, identified the allowed derivative and partial-compositeness couplings for each model class, and derived the characteristic decay patterns of the sextet hyperbaryons. The sextet states carry electric charges of $1/3$, $-2/3$, and $-5/3$, and their decays are typically mediated either by the colour-octet pNGB $\pi_8$, which is universally present in all model classes, or, in constructions of class C1, by the colour-sextet pNGB $\pi_6$ together with the singlet hyperbaryon $\hat Q_1$. The $\pi_8$-mediated channels then give rise to top-rich final states, while the $\pi_6$-mediated one yields a characteristic $bb + E_T^{\rm miss}$ signature as the $\hat{Q}_1$ states escape the detector as missing transverse energy. We have then investigated the sensitivity of existing LHC searches to these states. Since no dedicated search for colour-sextet fermions is currently available, we have reinterpreted the results of public ATLAS and CMS analyses designed for supersymmetric final states with high jet multiplicities, multiple $b$-jets and large missing transverse momentum. 

For the individual sextet components decaying through the $\pi_8$ pNGB, the strongest bound is obtained for the $\mathcal{Q}_6^{-5/3}$ state, which is excluded for masses up to $2.1$ ($2.4$)~TeV at leading order (with a representative multiplicative factor $K=3$ applied to the rate), slightly above the corresponding mass limits of $2.0$ ($2.3$)~TeV for the states carrying an electric charge of $+1/3$ and $-2/3$. This hierarchy is consistent with the larger visible jet multiplicity of the $\mathcal{Q}_6^{-5/3}$ decay chain. When the full sextet multiplet of five states is considered simultaneously, the enhanced total production rate strengthens the exclusion to $2.4$ ($2.6$)~TeV. We emphasise that these bounds are comparable to current constraints on colour-octet fermionic resonances. This is a non-trivial result, reflecting the interplay between the sizeable QCD production rate of each sextet component, the multiplicity of five nearly degenerate states in the full multiplet, and the different decay topologies entering the signal. 

In models of class C1, the $\pi_6$-mediated channel provides an important complementary signature. For sizeable mass splittings between the $\mathcal{Q}_6$ and $\hat{Q}_1$ hyperbaryons the current exclusions on the full multiplet reach $2.7$ ($2.9$)~TeV at leading order (with $K=3$), competitive with the $\pi_8$-mediated bounds. For more realistic compressed spectra, the $b$-jets become soft and the sensitivity of the available recasts is reduced, leaving this region as an important target for dedicated analyses. 

Conservative extrapolations to the HL-LHC luminosity of $3~\mathrm{ab}^{-1}$ at $\sqrt{s} = 13$~TeV indicate that the $\pi_8$-mediated reach for the full multiplet approaches $3$~TeV when including the representative $K$-factor of 3, and a similar reach is obtained in the $\pi_6$-mediated channel for favourable mass splittings. These results therefore provide a complementary picture of the most relevant low-energy sextet signatures in the benchmark classes considered here.

Our work has identified several directions that deserve further attention. A calculation of the next-to-leading-order QCD corrections to colour-sextet fermion pair production is the most pressing theoretical need, as it would substantially reduce the dominant uncertainty in the mass limits and remove the need for representative $K$-factors such as those used throughout this analysis. Second, dedicated LHC analyses targeting the specific topologies identified here, namely multi-top final states from the $\pi_8$ channel and $b\bar{b} + E_T^{\rm miss}$ signals from the $\pi_6$ channel, could improve significantly over recast bounds, particularly in the compressed regime where generic searches lose sensitivity. Third, single-production channels, although model-dependent, may become relevant for heavier sextets and should be studied systematically in the HL-LHC regime. Finally, the large QCD production rates associated with sextet colour representations make these states particularly compelling targets for future hadron colliders, such as a prospective FCC-hh operating at $\sqrt{s}=100$~TeV where a dedicated study could probe mass scales well beyond the reach of the LHC programme.

\section*{Acknowledgements}

This work has been partly supported by the PROCOPE project nr.~57755908 and 52970ZL, funded by the French Ministry for Europe and Foreign Affairs, the French Ministry for Higher Education and Research and the German Academic Exchange Service (DAAD). J.H.~and M.K.~have been supported by DFG research training group GRK 2994. R.C.~and W.P.~have been supported by DFG project PO1337/12-1. B.F. and  M.D.G. are supported in part by Grant ANR-21-CE31-0013, Project DMwithLLPatLHC, from the French \textit{Agence Nationale de la Recherche} (ANR). 

\appendix

\section{Calculation of relevant couplings}\label{app:calculation}
In this appendix, we collect the technical details underlying the baryonic operators and the associated interactions. We begin in \cref{app:embeddings} by specifying the CCWZ construction for the coloured sector and the embedding of the composite baryons into representations of the unbroken symmetry group. We next derive the corresponding interaction terms relevant for the sextet phenomenology, focusing both on derivative and partial compositeness couplings in \cref{app:derivative_int,app:pcint} respectively.

\subsection{CCWZ construction and field embeddings}\label{app:embeddings}
We start by detailing the CCWZ construction for the coloured sector and the embedding of the composite baryons into representations of the unbroken symmetry group. The coloured sector is based on different symmetry-breaking patterns depending on the model class, as shown in \cref{eq:gfbreaking}. In models of class C1, the coset is $\SU(6)/\SO(6)$, while models of class C2 are based on $\SU(6)/\Sp(6)$. Models of class C3 instead realise a $\SU(3)^2/\SU(3)$ structure. Despite these differences, a common matrix notation can be adopted to describe the vacuum alignment, which takes the block form
\begin{align}
    \Sigma_{\chi,0} =
    \begin{pmatrix}
        0 & \pm \mathds{1}_3 \\
        \mathds{1}_3 & 0
    \end{pmatrix}\,,
\end{align}
where the relative sign distinguishes between the orthogonal ($+$) and symplectic ($-$) realisations of the unbroken group. In class C3, a similar block form with the plus sign can be used as a convenient embedding of the $\SU(3)^2/\SU(3)$ structure, even though this coset 
is of complex type and does not correspond to an orthogonal or symplectic condensate.

The QCD subgroup $\SU(3)_c$ is embedded into the unbroken group through generators acting diagonally on the two $3\times3$ blocks. In the matrix notation introduced above, one may write
\begin{align}
    T^a =
    \frac{1}{\sqrt{2}}
    \begin{pmatrix}
        \frac{1}{2}\lambda^a & 0 \\
        0 & -\frac{1}{2}(\lambda^a)^T
    \end{pmatrix},
    \qquad a=1,\dots,8\,,
\end{align}
where $\lambda^a$ are the Gell-Mann matrices and we recall that the lower block transforms in the conjugate representation $\bar{\mathbf{3}}$. These generators satisfy the standard $\SU(3)$ algebra and act on the pNGB and baryon matrices. The Abelian generator $T_X$ is then embedded as
\begin{align}
    T_X = X_\chi
    \begin{pmatrix}
        \mathds{1}_3 & 0 \\
        0 & -\mathds{1}_3
    \end{pmatrix}\,,
\end{align}
where $X_\chi=-1/3$ is the $\U(1)_X$ charge assigned to the $\chi$ hyperfermion. This choice ensures that the pNGB and baryon multiplets carry the $X$-charges quoted in \cref{sec:models}, and that the conventional hypercharge is recovered through the usual relation $Y = T_R^3 + X$. For the $\SU(6)/\SO(6)$ and $\SU(6)/\Sp(6)$ cosets relevant to constructions of classes C1 and C2, the generators $T^a$ and $T_X$ belong to the unbroken subgroup and therefore satisfy the vacuum-invariance condition
\begin{align}
    T\,\Sigma_{\chi,0}+\Sigma_{\chi,0}\,T^T=0\,.
\end{align}
In the matrix embedding adopted for models of class C3, the diagonal $\SU(3)_c$ generators satisfy the same relation, although in that case it should be understood as a property of the chosen $6\times6$ representation of the $\SU(3)^2/\SU(3)$ structure.

With these conventions for the vacuum alignment and the embedding of the unbroken generators, the coloured pNGBs can be assembled into a matrix $\Pi_\chi$ transforming non-linearly under the global symmetry. In all cases, it is convenient to use a common block form,
\begin{align}\renewcommand{\arraystretch}{1.4}
    \Pi_\chi =
    \begin{pmatrix}
        \frac{1}{\sqrt{2}}\,\pi_8 & \Phi_\chi \\
        \Phi_\chi^\dagger & \frac{1}{\sqrt{2}}\,\pi_8^T
    \end{pmatrix}\,,
\end{align}
where $\pi_8$ denotes the colour-octet pNGB written as a traceless $3\times3$ matrix, while the off-diagonal block $\Phi_\chi$ collects the additional coloured pNGBs that depend on the model class. In class C1, the off-diagonal block contains the colour-sextet pNGB and its conjugate ($\Phi_\chi\equiv \pi_6$ and $\Phi_\chi^\dagger \equiv \pi_6^c$), while in class C2 it contains the colour-triplet pNGB and its conjugate ($\Phi_\chi \equiv \pi_3^\dagger$ and $\Phi_\chi^\dagger \equiv \pi_3$). In class C3, no additional coloured pNGBs are present, and only the octet survives. The non-linear realisation of the global symmetry is then implemented in the standard CCWZ manner~\cite{Coleman:1969sm, Callan:1969sn}, through the Goldstone matrix
\begin{align}
    U_\chi = \exp\!\left(\frac{2i}{f_\chi}\,\Pi_\chi\right)\,,
\end{align}
where $f_\chi$ denotes the decay constant associated with the coloured pNGB sector. The CCWZ symbol $d_\mu$ relevant for the derivative interactions is subsequently defined as
\begin{align}
    d_\mu = 2i\,\mathrm{Tr}\!\left(U_\chi^\dagger D_\mu U_\chi\, X^I\right)X^I
    = -\frac{2}{f_\chi}\,\partial_\mu \Pi_\chi + \cdots\,,
\end{align}
where $X$ denotes the broken generators. At leading order in the pNGB fields, this object provides the universal building block for the derivative couplings among the baryons and the coloured pNGBs. It transforms as a vector in the coset representation under the unbroken group, and therefore lies in the $\mathbf{20}$ representation of $\SO(6)$ in models of class C1 and in the $\mathbf{14}$ representation of $\Sp(6)$ for models of the class C2. In scenarios of class C3, the broken generators (and therefore $d_\mu$) transform as an octet of the diagonal $\SU(3)_c$. This observation immediately constrains which baryon bilinears can couple through $d_\mu$, and therefore which derivative interactions are allowed in each model class, as detailed in \cref{app:derivative_int}. 

We next specify the embedding of the composite baryons into matrix representations of the unbroken group. For models of class C1, the relevant baryons transform in the $\mathbf{20}$, $\mathbf{15}$ and $\mathbf{1}$ representations of $\SO(6)$. They can be embedded as
\begin{align}\label{eq:embeddings_c1}\renewcommand{\arraystretch}{1.4}
    \mathcal{B}_{\mathbf{20}} =
    \begin{pmatrix}
        -\mathcal{Q}_6 & \frac{1}{\sqrt{2}}\,\mathcal{Q}_8 \\
        \frac{1}{\sqrt{2}}\,\mathcal{Q}_8^T & -\mathcal{Q}_6^c
    \end{pmatrix}\,,\qquad 
    \mathcal{B}_{\mathbf{15}} =
    \begin{pmatrix}
        -Q_3^c & -\frac{1}{\sqrt{2}}\,Q_8 - Q_1 \\
        \frac{1}{\sqrt{2}}\,Q_8^T + Q_1^T & -Q_3
    \end{pmatrix}\,,\qquad
    \mathcal{B}_{\mathbf{1}} = \hat Q_1\,\Sigma_{\chi,0}\,.
\end{align}
For models of class C2, the baryons instead transform in the $\mathbf{21}$, $\mathbf{14}$ and $\mathbf{1}$ representations of $\Sp(6)$, with embeddings
\begin{align}\renewcommand{\arraystretch}{1.4}
    \mathcal{B}_{\mathbf{21}} =
    \begin{pmatrix}
        -\mathcal{Q}_6 & \frac{1}{\sqrt{2}}\,\mathcal{Q}_8 + \mathcal{Q}_1 \\
        \frac{1}{\sqrt{2}}\,\mathcal{Q}_8^T + \mathcal{Q}_1^T & -\mathcal{Q}_6^c
    \end{pmatrix}\,,\quad\
    \mathcal{B}_{\mathbf{14}} =
    \begin{pmatrix}
        -Q_3^c & -\frac{1}{\sqrt{2}}\,Q_8 \\
        \frac{1}{\sqrt{2}}\,Q_8^T & -Q_3
    \end{pmatrix}\,,\quad \
    \mathcal{B}_{\mathbf{1}} = \hat Q_1\,\Sigma_{\chi,0}\,.\qquad
\end{align}
Finally, for models of class C3, the baryons arise from two distinct classes of gauge-invariant operators, and they can be conveniently embedded in the same $6\times6$ matrix notation. Following the conventions of Appendix~A.1 of~\cite{Caliri:2024jdk}, the relevant embeddings are
\begin{align}\renewcommand{\arraystretch}{1.4}
    \mathcal{B}^{F,c}_{\mathbf{3}} =
    \begin{pmatrix}
        -Q_3^c & 0 \\
        0 & 0
    \end{pmatrix}\,,\qquad
    \mathcal{B}^{B,c}_{\mathbf{3}} =
    \begin{pmatrix}
        0 & -Q_3^c \\
        0 & 0
    \end{pmatrix}\,,\qquad
    \mathcal{B}^{B}_{\mathbf{6}} =
    \begin{pmatrix}
        0 & \mathcal{Q}_6 \\
        0 & 0
    \end{pmatrix}\,,
\end{align}
where the superscripts $F$ and $B$ label the two inequivalent embedding structures entering the partial compositeness construction, corresponding respectively to (diagonal) fundamental-like and (off-diagonal) bi-fundamental-like operators in the matrix realisation.

\subsection{Derivative interactions}\label{app:derivative_int}
The derivative interactions among the baryons and the coloured pNGBs are obtained by contracting the CCWZ symbol $d_\mu$ with the baryon embeddings introduced above. We recall that $d_\mu$ transforms in the same representation of the unbroken group as the broken generators and that only those bilinears $\bar{\mathcal B}_i\, d_\mu\, \mathcal B_j$ that contain an invariant singlet are allowed. At leading order in the pNGB fields, the derivative interactions thus take the schematic form
\begin{align}\label{eq:lder_general}
    \mathcal{L}_{\rm der} \supset 
    c_{ij}\,
    \bar{\mathcal B}_i\,\bar\sigma^\mu d_\mu\,\mathcal B_j
    + \mathrm{H.c.}\,,
\end{align}
where the allowed contractions depend on the representations of the baryon multiplets under the unbroken group. Let us however note that since $d_\mu$ acts on a two-index representation in \cref{eq:lder_general}, we need in principle to calculate
\begin{align}
    \bar{\mathcal B}_i \bar\sigma^\mu d_\mu \mathcal B_j = d_\mu^I \Tr(\bar{\mathcal B}_i \bar\sigma^\mu X^I \mathcal B_j + \bar{\mathcal B}_i \bar\sigma^\mu \mathcal B_j (X^I)^T)\,.
\end{align}
The embedding of $\mathcal B_{20}$ (which contains the sextet fields) then corresponds to the broken generators up to the vacuum, $\mathcal B_{20} = \mathcal B_{20}^I X^I \Sigma_0$, whereas for the other representation we have $\mathcal B_{15} = \mathcal B_{15}^a T^a \Sigma_0$. The phenomenologically relevant combinations can then all be written as 
\begin{align}
    d_\mu^I \Tr(\bar{\mathcal B}_i \bar\sigma^\mu  \mathcal B_{20}^J  X^I X^J \Sigma_0 + \bar{\mathcal B}_i \bar\sigma^\mu \mathcal B_{20}^J X^J \Sigma_0 (X^I)^T)\,.
\end{align}
Using $X^I \Sigma_0 - \Sigma_0 (X^I)^T=0$, this simplifies to
\begin{align}
    \Tr(\bar{\mathcal B}_i \bar\sigma^\mu   d_\mu^I X^I \mathcal B_{20} \Sigma_0)\,,
\end{align}
so it is sufficient to calculate one single combination. In the following, only the couplings yielding interaction terms relevant for the sextet phenomenology are retained, and we derive them explicitly from the matrix embeddings discussed in \cref{app:embeddings}. 

We recall that in models of class C1, the broken generators of the $\SU(6)/\SO(6)$ coset form a $\mathbf{20}$ of $\SO(6)$, and the CCWZ symbol $d_\mu$ transforms in the same representation. The possible invariant contractions are thus
\begin{align}\label{eq:lderc1}
    \mathcal L_\mathrm{der}^{\rm C1} =
    c_{20}\,\overline{\mathcal B}_{\mathbf{20}}\,\bar\sigma^\mu d_\mu\,\mathcal B_{\mathbf{20}}
    + c_{15}\,\overline{\mathcal B}_{\mathbf{15}}\,\bar\sigma^\mu d_\mu\,\mathcal B_{\mathbf{20}}
    + c_1\,\overline{\mathcal B}_{\mathbf{1}}\,\bar\sigma^\mu d_\mu\,\mathcal B_{\mathbf{20}}
    + \mathrm{H.c.}\,.
\end{align}
The first term couples one $\mathbf{20}$ multiplet to itself and does not induce decays of the sextet baryons, while other self-couplings of the $\mathcal{B}_{\mathbf{15}}$ are also present but not relevant as they do not involve sextet baryons. We therefore focus on the second and third contractions, which, once expanded to first order in the pNGB fields, read
\begin{equation}\begin{split}
     &\Tr \overline{\mathcal B}_{\mathbf{1}} \,\bar{\sigma}^\mu \partial_\mu \Pi_\chi \,\mathcal B_{\mathbf{20}} = 
       \frac{1}{2}\,\bar{\hat Q}_1 \,\bar{\sigma}^\mu \partial_\mu \pi_{6}^c \,\mathcal{Q}_{6} 
       + \frac{1}{2}\,\bar{\hat Q}_1 \,\bar{\sigma}^\mu \partial_\mu \pi_{6} \,\mathcal{Q}_{6}^c
       + \frac{1}{2}\,\bar{\hat Q}_1 \,\bar{\sigma}^\mu \partial_\mu \pi_8 \,\mathcal{Q}_8\,,\\
    &\Tr \overline{\mathcal B}_{\mathbf{15}} \,\bar{\sigma}^\mu \partial_\mu \Pi_\chi \,\mathcal B_{\mathbf{20}} =
      \frac{1}{2}\,\bar Q_1 \,\bar{\sigma}^\mu \partial_\mu \pi_{6} \,\mathcal{Q}_{6}^c
     - \frac{1}{2}\,\bar Q_1 \,\bar{\sigma}^\mu \partial_\mu \pi_{6}^c \,\mathcal{Q}_6
     + \frac{1}{2}\,\bar Q_1 \,\bar{\sigma}^\mu \partial_\mu \pi_8 \,\mathcal{Q}_8\\ &\hspace{2cm}
     - \frac{i}{2}\,\bar Q_3^c \,\bar{\sigma}^\mu \partial_\mu \pi_8 \,\mathcal{Q}_6
     - \frac{i}{2}\,\bar Q_3 \,\bar{\sigma}^\mu \partial_\mu \pi_8 \,\mathcal{Q}_{6}^c
     + \frac{i}{2}\,\bar Q_3^c \,\bar{\sigma}^\mu \partial_\mu \pi_6 \,\mathcal{Q}_8
     + \frac{i}{2}\,\bar Q_3 \,\bar{\sigma}^\mu \partial_\mu \pi_{6}^c \,\mathcal{Q}_8
     \\ &\hspace{2cm}
     + \frac{1}{2\sqrt{2}}\,\bar Q_8 \,\bar{\sigma}^\mu \partial_\mu \pi_6 \,\mathcal{Q}_{6}^c
     - \frac{1}{2\sqrt{2}}\,\bar Q_8 \,\bar{\sigma}^\mu \partial_\mu \pi_{6}^c \,\mathcal{Q}_{6}
     + \frac{i}{4\sqrt{2}}\,\bar Q_8^a \,\bar{\sigma}^\mu \partial_\mu \pi_8^b \,\mathcal{Q}_8^c f^{abc}\,.
\end{split}\end{equation}
For compactness, colour indices are suppressed whenever the invariant contraction is unique. Explicitly, one has
\begin{equation}\begin{split}
    \pi_6^c \mathcal Q_6 = \pi_{6,s}^c \mathcal Q_6^s,\qquad
    \pi_8 \mathcal Q_8 = \pi_8^a \mathcal Q_8^a,\qquad
    Q_8 \pi_6 \mathcal Q_6^c = Q_8^a \pi_6^s \mathcal Q_{6,t}^c [t_6^a]_s{}^t,\\[.1cm]
    Q_3 \pi_8 \mathcal Q_6 = Q_3^i \pi_8^a \mathcal Q_6^s J_{sia},\qquad
    Q_3^c \pi_8 \mathcal Q_6^c = Q_{3,i}^c \pi_8^a \mathcal Q_{6,s}^c \bar J^{sia}\,,
\end{split}\end{equation}
where $t_6^a$ are the generators in the sextet representation, and $J_{sia}$ and $\bar J^{sia}$ are the Clebsch-Gordan coefficients for the $\mathbf{3}\otimes\mathbf{6}\otimes\mathbf{8}$ contraction following the conventions of~\cite{Carpenter:2021rkl}.

In scenarios of class C2, $d_\mu$ transforms in the  $\mathbf{14}$ of $\Sp(6)$, consistently with the coloured pNGB. The leading derivative interactions relevant for the sextet baryons are therefore
\begin{align}
    \mathcal L_\mathrm{der}^{\rm C2} =
    c_{14}\,\overline{\mathcal B}_{\mathbf{14}}\,\bar\sigma^\mu d_\mu\,\mathcal B_{\mathbf{21}}
    + c_{21}\,\overline{\mathcal B}_{\mathbf{14}}\,\bar\sigma^\mu d_\mu\,\mathcal B_{\mathbf{14}}
    + \mathrm{H.c.}\,,
\end{align}
where the second term is a self-coupling of $\mathcal{B}_{\mathbf{14}}$ and does thus not induce any decay. Therefore, only the first contraction generates sextet interactions and is considered further. Expanding the trace to first order in the pNGB fields yields
\begin{equation}\begin{split}
    & \Tr \overline{\mathcal B}_{\mathbf{14}} \,\bar{\sigma}^\mu \partial_\mu \Pi_\chi \,\mathcal B_{\mathbf{21}} = 
    \frac{1}{2}\,\bar Q_3^c \,\bar{\sigma}^\mu \partial_\mu \pi_3 \,\mathcal Q_1
     - \frac{1}{2}\,\bar Q_3 \,\bar{\sigma}^\mu \partial_\mu \pi_3^c \,\mathcal Q_1
     - \frac{i}{2}\,\bar Q_3 \,\bar{\sigma}^\mu \partial_\mu \pi_8 \,\mathcal Q_6\\ &\qquad
     - \frac{i}{2}\,\bar Q_3^c \,\bar{\sigma}^\mu \partial_\mu \pi_8 \,\mathcal Q_6^c
     + \frac{i}{2}\,\bar Q_8 \,\bar{\sigma}^\mu \partial_\mu \pi_3 \,\mathcal Q_6
     + \frac{i}{2}\,\bar Q_8 \,\bar{\sigma}^\mu \partial_\mu \pi_3^c \,\mathcal Q_6^c
     + \frac{1}{2\sqrt{2}}\,\bar Q_3 \,\bar{\sigma}^\mu \partial_\mu \pi_3^c \,\mathcal Q_8\\ &\qquad
     + \frac{1}{2\sqrt{2}}\,\bar Q_3^c \,\bar{\sigma}^\mu \partial_\mu \pi_3 \,\mathcal Q_8
     + \frac{i}{4\sqrt{2}}\,\bar Q_8^a \,\bar{\sigma}^\mu \partial_\mu \pi_8^b \,\mathcal Q_8^c f^{abc}\,,
\end{split}\end{equation}
where as in the C1 case, we suppress colour indices whenever the invariant contraction is unique. 

Finally, in models of class C3, the CCWZ symbol $d_\mu$ transforms as an octet of the diagonal $\SU(3)_c$, and the derivative interactions relevant for the sextet baryons reduce to the unique $\mathbf{3}\otimes\mathbf{6}\otimes\mathbf{8}$ invariant. Using the embeddings introduced above, one immediately finds
\begin{align}
    \mathcal L_\mathrm{der}^{\rm C3} \supset 
    c_1 \,\bar Q_3\,\bar\sigma^\mu \partial_\mu \pi_8\,\mathcal Q_6^c
    + c_2\,\bar Q_3^c\,\bar\sigma^\mu \partial_\mu \pi_8\,\mathcal Q_6
    + \mathrm{H.c.}
\end{align}

\subsection{Partial compositeness interactions}\label{app:pcint}
We now turn to the interactions induced by partial compositeness which arise from the linear mixing of the elementary SM quarks with composite baryonic operators. In contrast to the derivative couplings discussed in \cref{app:derivative_int}, these interactions depend on the embedding of the elementary fermions into representations of the global flavour symmetry. Their generic structure is given by \cref{eq:PC_general},
\begin{align}
    \mathcal{L}_{\rm PC} \supset  -\lambda_L\,  q_L\, \mathcal O_L - \lambda_R\,  t_R^c\, \mathcal O_R    + \mathrm{H.c.}\,,
\end{align}
where the composite operators $\mathcal O_{L,R}$ are built from the baryonic multiplets dressed by the pNGB matrices through the CCWZ construction, and the $\lambda_L$ and $\lambda_R$ parameters denote generic couplings involving the left-handed and right-handed SM quarks respectively. For the electroweak sector, the corresponding Goldstone matrix is defined by
\begin{align}
    U_\psi = \exp\!\left(\frac{i\sqrt{2}}{f_\psi}\,\Pi_\psi\right)\,,
\end{align}
with the associated pNGB matrix being denoted by $\Pi_\psi$ in complete analogy with the coloured-sector construction based on $U_\chi$. However, in the rest of this section, only the leading terms in the expansion in $\Pi_\chi$ are retained explicitly, while the dependence on $\Pi_\psi$ is kept implicit unless needed. Expanding these terms to first order in the coloured pNGBs then generates interactions between the sextet baryons, the pNGBs and the SM quarks whose precise form depends on the embedding of the elementary fermions into the global flavour symmetry. 

In models of class C1, the elementary quarks can be embedded into different representations of the global $\SU(6)\times\SU(5)$ flavour symmetry, leading to inequivalent partial compositeness structures. For the left-handed doublet $q_L$, the two relevant embeddings are the antisymmetric one $q_L\in (\overline{\mathbf{15}},\bar{\mathbf{5}})$ (or $(\mathbf{15}, \bar{\mathbf{5}})$ which yields the same phenomenology), and the adjoint one $q_L\in (\mathbf{35},\mathbf{5})$. For the antisymmetric embedding, the partial compositeness interaction can be written as
\begin{align}
   \mathcal{L}_\mathrm{PC}^{\rm C1,\bar A} = \lambda_L\, \zeta_{L,\bar A} \cdot U_\chi U_\psi\, \mathcal{B}_{\mathbf{15}}\, U_\chi^T + \mathrm{H.c.}\,,
\end{align}
in which
\begin{align}\label{eq:zetaA}
    \zeta_{L,\bar A} = \begin{pmatrix} \frac12 \epsilon_{ijk}\zeta_{L,k} & 0 \\ 0 & 0 \end{pmatrix}
    \qquad\text{with}\qquad
    \zeta_L = (b_L,-t_L,0,0,0)\,.
\end{align}
Here the Levi-Civita tensor $\epsilon_{ijk}$ is used to construct the antisymmetric $3\times3$ block entering the $\overline{\mathbf{15}}$ embedding of $\SU(6)$: the electroweak spurion content is carried by the five-plet $\zeta_L$ of $\SU(5)$ while the $\epsilon_{ijk}$ contraction fixes the antisymmetric structure in colour space. Expanding to first order in the coloured pNGBs and keeping the terms relevant for the baryonic spectrum yields
\begin{align}
    \mathcal{L}_\mathrm{PC}^{\rm C1,\bar A} \supset \frac{\lambda_L}{f_\chi}\,\zeta_L 
       \left(2\,\pi_6\,Q_8 -\sqrt{2}\,i\,\pi_8\,Q_3^c \right) + \mathrm{H.c.}
\end{align}
No direct coupling involving the sextet baryon $\mathcal Q_6$ is thus generated at this order. For the adjoint embedding, the corresponding interaction reads
\begin{align}
    \mathcal{L}_\mathrm{PC}^{\rm C1,D} = \zeta_{L,D} \cdot U_\chi \left( \lambda_L\,\widetilde{\mathcal{B}}_{\mathbf{15}} + \lambda_L'\,\widetilde{\mathcal{B}}_{\mathbf{20}} \right) U_\psi^\dagger U_\chi^\dagger
    + \mathrm{H.c.}\,,
\end{align}
where
\begin{align}\label{eq:zetaD}
    \zeta_{L,D} = \begin{pmatrix} 0 & 0 \\ \frac{1}{2}\epsilon_{ijk}\zeta_{L,k} & 0 \end{pmatrix}
    \qquad\text{with}\qquad
    \zeta_L = (0,0,t_L,b_L,0)\,.
\end{align}
The $\widetilde{\mathcal{B}}_{i}$ embeddings corresponds to those in \cref{eq:embeddings_c1} up to an additional rotation permuting the diagonal and off-diagonal elements of the hyperbaryon matrices relevant for the adjoint embedding. Expanding again to first order in the coloured pNGBs, one obtains
\begin{align}
    \mathcal{L}_\mathrm{PC}^{\rm C1,D} \supset \frac{\lambda_L}{2}\,\zeta_L\,Q_3^c 
      + \frac{2\lambda_L'}{f_\chi}\,\zeta_L \left( \pi_8\,\mathcal Q_6 - \pi_6\,\mathcal Q_8 \right)
      + \mathrm{H.c.}
\end{align}
The first term gives rise to the usual linear mixing of the elementary doublet with the top-partner multiplet, while the second term contains the direct $\mathcal Q_6$-$\pi_8$-$q_L$ interaction relevant for the sextet decay $\mathcal Q_6\to \pi_8\,\bar t_L$. For the right-handed top quark, one analogously introduces the spurion 
\begin{align}
    \zeta_R^c = (0,0,0,0,\mp i t_R^c)\,,
\end{align}
where the minus sign corresponds to the antisymmetric embedding and the plus sign to the adjoint one. As in the left-handed case, the antisymmetric embedding does not generate any direct coupling to the sextet baryon at leading order, while the adjoint embedding yields
\begin{align}
    \mathcal{L}_\mathrm{PC}^{\rm C1,R} \supset 
     \frac{\lambda_R}{f_\chi}\,\zeta_{R}^c\,\pi_8\,\mathcal Q_{6}^c
    + \mathrm{H.c.}\,,
\end{align}
which provides the complementary decay mode $\mathcal Q_6\to \pi_8\,\bar t_R$. Therefore, in models of class C1, only the adjoint embedding generates direct partial-compositeness couplings of the sextet baryons to SM quarks at leading order.

In models of class C2, the partial compositeness mechanism still follows the generic structure of \cref{eq:PC_general}, but the relevant composite operators must now be embedded into representations of $\SU(6)\times\SU(5)$ compatible with the $\SU(6)/\Sp(6)$ coset. As in class C1, the left-handed doublet $q_L$ admits two inequivalent embeddings, namely an antisymmetric one $q_L\in (\overline{\mathbf{15}},\bar{\mathbf{5}})$ and an adjoint one $q_L\in (\mathbf{35},\mathbf{5})$. For the antisymmetric embedding, the partial compositeness interaction can be written as
\begin{align}
   \mathcal{L}_\mathrm{PC}^{\rm C2,\bar A} = \zeta_{L,\bar A}\cdot U_\chi U_\psi
      \left(\lambda_L\,\mathcal B_{\mathbf{14}}+\lambda_L'\,\mathcal B_{\mathbf{1}}\right) U_\chi^T
      + \mathrm{H.c.}\,,
\end{align}
where the spurion embeddings are the same as in the C1 case. Expanding to first order in the coloured pNGBs then yields
\begin{align}
    \mathcal{L}_\mathrm{PC}^{\rm C2,\bar A} \supset 
      -\lambda_L\,\frac{\sqrt2\,i}{f_\chi}\,\zeta_L \left(\pi_3^c\,Q_8+\pi_8\,Q_3^c\right)
    + \lambda_L'\,\frac{2i}{f_\chi}\,\zeta_L\,\pi_3^c\,\hat Q_1
    + \mathrm{H.c.}
\end{align}
Once again, no direct coupling involving the sextet baryon $\mathcal Q_6$ is generated at this order. For the adjoint embedding, the corresponding interaction reads
\begin{align}
    \mathcal{L}_\mathrm{PC}^{\rm C2,D} = \zeta_{L,D}\cdot U_\chi
      \left(\lambda_L\,\widetilde{\mathcal B}_{\mathbf{14}} + \lambda_L'\, \widetilde{\mathcal B}_{\mathbf{21}}\right)   U_\psi^\dagger U_\chi^\dagger
    + \mathrm{H.c.}\,,
\end{align} 
which gives, after expanding again to first order in the coloured pNGBs, 
\begin{align}
    \mathcal{L}_\mathrm{PC}^{\rm C2,D} \supset 
       \frac{\lambda_L}{2}\,\zeta_L\,Q_3^c  + \frac{\lambda_L'}{f_\chi}\,\zeta_L
          \left( 2\,\pi_8\,\mathcal Q_6 + 2i\,\pi_3^c\,\mathcal Q_1 - \sqrt2\,i\,\pi_3^c\,\mathcal Q_8 \right)
    + \mathrm{H.c.}
\end{align}
The first term corresponds to the usual linear mixing with the top-partner multiplet, while the second term contains the direct $\mathcal Q_6$-$\pi_8$-$q_L$ interaction relevant for the sextet phenomenology. Moreover, the additional terms involving the colour-triplet pNGB $\pi_3$ only couple to the non-sextet baryons. Similarly, for the right-handed top quark, the antisymmetric embedding does not generate any direct coupling to the sextet baryon at leading order, while the adjoint embedding yields
\begin{align}
    \mathcal{L}_\mathrm{PC}^{\rm C2,R}\supset \frac{\lambda_R}{f_\chi}\,\zeta_{R}^c\,\pi_8\,\mathcal Q_{6}^c\,   + \mathrm{H.c.}
\end{align}
Therefore, also in models of class C2, only the adjoint embedding generates direct partial-composite\-ness couplings of the sextet baryons to SM quarks at leading order, with the role of the additional coloured pNGB now played by the triplet $\pi_3$ whose couplings nevertheless involve only the non-sextet baryon multiplets. 

In models of class C3, while the structure of the partial compositeness interactions differs from that of models of classes C1 and C2, the relevant composite operators also fall into two inequivalent classes. In the naming scheme introduced at the end of \cref{app:embeddings}, these are represented by diagonal fundamental-like ($\zeta_{L,F} \equiv \zeta_{L,\bar A}$ in the notation of \cref{eq:zetaA}) and off-diagonal bi-fundamental-like ($\zeta_{L,B} \equiv \zeta_{L,D}$ in the notation of \cref{eq:zetaD}) embeddings For the left-handed doublet $q_L$, the first possibility is associated with the operator $\chi\psi\chi$,
\begin{align}
    \mathcal{L}_\mathrm{PC}^{\rm C3,F} = \lambda_L\,\zeta_{L,F}\cdot U_\chi U_\psi\,\mathcal B_{\mathbf 3}^{F,c}\,U_\chi^T + \mathrm{H.c.}\,,
\end{align}
which can be written, after an expansion to first order in the coloured pNGBs, as
\begin{align}
    \mathcal{L}_\mathrm{PC}^{\rm C3,F} \supset \frac{\lambda_L}{2}\,\zeta_L\,Q_3^c - \lambda_L\,\frac{\sqrt2\,i}{f_\chi}\,\zeta_L\,\pi_8\,Q_3^c  + \mathrm{H.c.}
\end{align}
Thus, as in the antisymmetric embeddings of mdoels of classes C1 and C2, no direct coupling involving the sextet baryon $\mathcal Q_6$ is generated at this order. The second possibility is associated with the operator $\chi\bar\psi\bar{\tilde\chi}$, 
\begin{align}
    \mathcal{L}_\mathrm{PC}^{\rm C3,B}  = \zeta_{L,B}\cdot U_\chi \left(
        \lambda_L\,\mathcal B_{\mathbf 3}^{B,c}  + \lambda_L'\,\mathcal B_{\mathbf 6}^{B} \right) U_\psi^\dagger U_\chi^\dagger + \mathrm{H.c.}
\end{align}
Expanding again to first order in the coloured pNGBs, one obtains
\begin{align}
    \mathcal{L}_\mathrm{PC}^{\rm C3,B} \supset \frac{\lambda_L}{2}\,\zeta_L\,Q_3^c - \frac{2\lambda_L'}{f_\chi}\,\zeta_L\,\pi_8\,\mathcal Q_6 + \mathrm{H.c.}\,,
\end{align}
which generates the direct $\mathcal Q_6$-$\pi_8$-$q_L$ interaction relevant for the sextet phenomenology. For the right-handed top quark, the same distinction applies: the operator class associated with $\tilde\chi\psi\tilde\chi$ plays the same role as the diagonal embedding above and does not generate a direct sextet coupling at leading order, while the operator class associated with $\bar\chi\bar\psi\tilde\chi$ yields
\begin{align}
    \mathcal{L}_\mathrm{PC}^{\rm C3,R} \supset \frac{\lambda_R}{f_\chi}\,\zeta_{R}^c\,\pi_8\,\mathcal Q_{6}^c  + \mathrm{H.c.}
\end{align}
Therefore, also in class C3, only one of the two operator classes generates direct partial-compositeness couplings of the sextet baryons to SM quarks at leading order. In contrast to classes C1 and C2, however, the distinction is not tied to two different embeddings of the same operator, but rather to the two distinct classes of gauge-invariant baryonic operators allowed by the complex representation structure of the hyperfermions.

At the level relevant for the sextet phenomenology, all model classes therefore admit direct sextet couplings of the form of \cref{eq:PC_sextet}
\begin{align}
    \mathcal L_{\rm PC} \supset 
    \frac{\lambda_L}{f_\chi}\,\zeta_L^i\,\pi_8^a\,\mathcal Q_6^s J_{sia}
    + \frac{\lambda_R}{f_\chi}\,\zeta_{R,i}^c\,\pi_8^a\,\mathcal Q_{6,s}^c \bar J^{sia}
    + \mathrm{H.c.}\,,
\end{align}
where all colour indices are now explicitly indicated. These interactions provide direct and unsuppressed decay channels $\mathcal Q_6\to \pi_8\,\bar t_{L,R}$, in contrast to the derivative couplings discussed in \cref{app:derivative_int} which instead proceed through baryon-to-baryon transitions.

\section{Details of the LHC recasts and their HL-LHC extrapolations}\label{app:HLLHC}
To estimate the projected sensitivity at the high-luminosity phase of the LHC, we extrapolate the limits of the considered analyses to an integrated luminosity of $3~\mathrm{ab}^{-1}$. While such a reprocessing is available in \madanalysis for some time~\cite{Araz:2019otb}, we implement the extrapolation using the scripts provided with \hackanalysis~\cite{Goodsell:2024aig}. These scripts read the \madanalysis cutflows and perform the statistical interpretation using \spey~\cite{Araz:2023bwx} for simplified likelihoods with Poisson-distributed bins and a multivariate Gaussian treatment of correlated uncertainties, and \pyhf~\cite{Heinrich:2021gyp} whenever a corresponding statistical model is available. 

We adopt a conservative systematic-dominated prescription in which the signal and background yields are rescaled linearly with the integrated luminosity. The absolute background uncertainties are also taken to scale linearly with luminosity, corresponding to fixed fractional systematic uncertainties. Equivalently, the elements of the covariance matrix scale quadratically with the luminosity. This choice is conservative compared with a statistics-dominated extrapolation in which the absolute uncertainties would scale only as the square root of the luminosity. For full \pyhf models, such a rescaling is in general not straightforward because the likelihood may contain several nuisance parameters, constraints and correlations whose luminosity dependence cannot be inferred unambiguously. However, for the \texttt{ATLAS-SUSY-2018-31} analysis which provides the dominant (ATLAS) limit in most of the cases considered in our study, simplified \pyhf models are available following the construction of~\cite{ATL-PHYS-PUB-2021-038}. These models consist of binned Poisson likelihoods with background expectations and uncertainties specified in a simplified form, with correlations approximated by a single Gaussian constraint. This structure allows for a controlled luminosity extrapolation by rescaling the signal and background yields, together with the corresponding absolute uncertainties, according to the prescription described above.

We stress that this extrapolation is performed at a fixed centre-of-mass energy of $\sqrt{s}=13$~TeV. A more realistic HL-LHC projection at $\sqrt{s}=14$~TeV would require instead a dedicated simulation of both the signal and background samples, including their different scaling with energy after the analysis selections. Since this effect is not included and since the uncertainties are assumed to remain systematics dominated, the projected limits should be interpreted as conservative estimates of the HL-LHC reach.

\begin{figure}[t]
    \centering
    \includegraphics[width=0.45\linewidth]{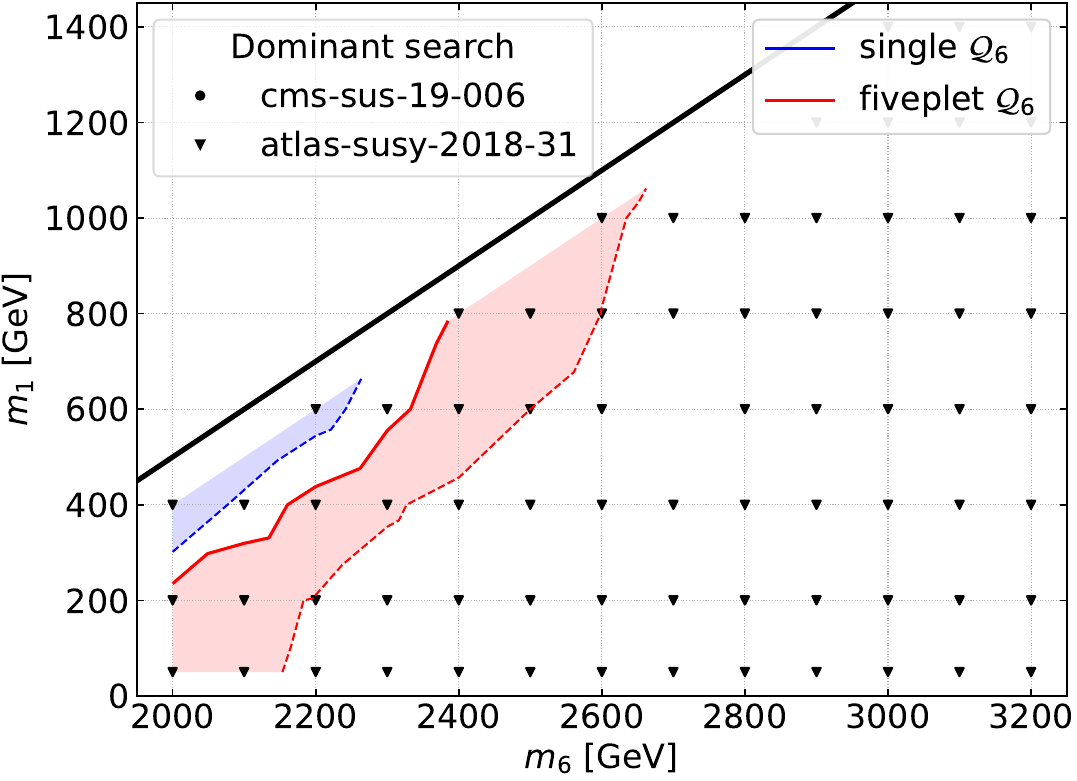}
    \includegraphics[width=0.45\linewidth]{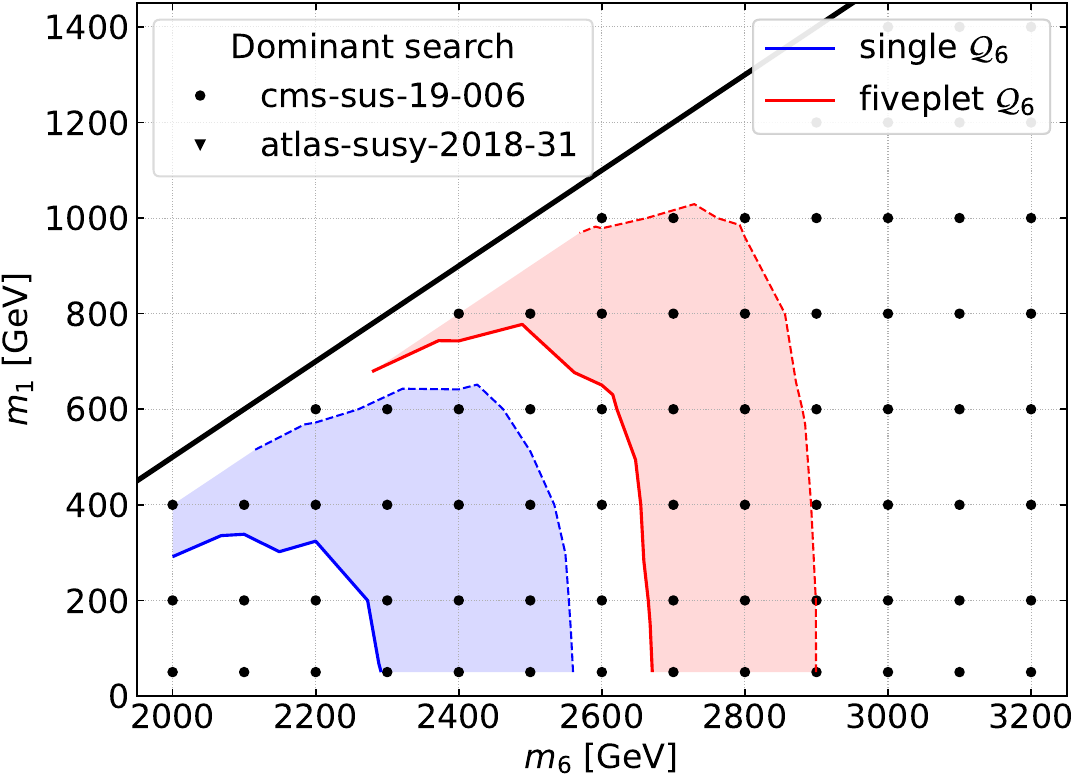}
    \caption{Separate LHC bounds on pair-produced sextet hyperbaryons decaying through the class-C1 channel $\mathcal Q_6\to\pi_6\hat Q_1\to bb+E_T^{\rm miss}$, assuming a $100\%$ branching ratio. We present predictions for the \texttt{ATLAS-SUSY-2018-31} (left) and \texttt{CMS-SUS-19-006} (right) analyses. The exclusions are shown in the $(m_6,m_1)$ plane, both for a single $\mathcal Q_6$ state (blue) and for the full five-component multiplet (red). Solid contours correspond to leading-order sextet pair production, while dashed contours include a representative $K$-factor of $3$.\label{fig:q6_DM_exclusion_details}} \vspace{.4cm}
    \includegraphics[width=0.45\linewidth]{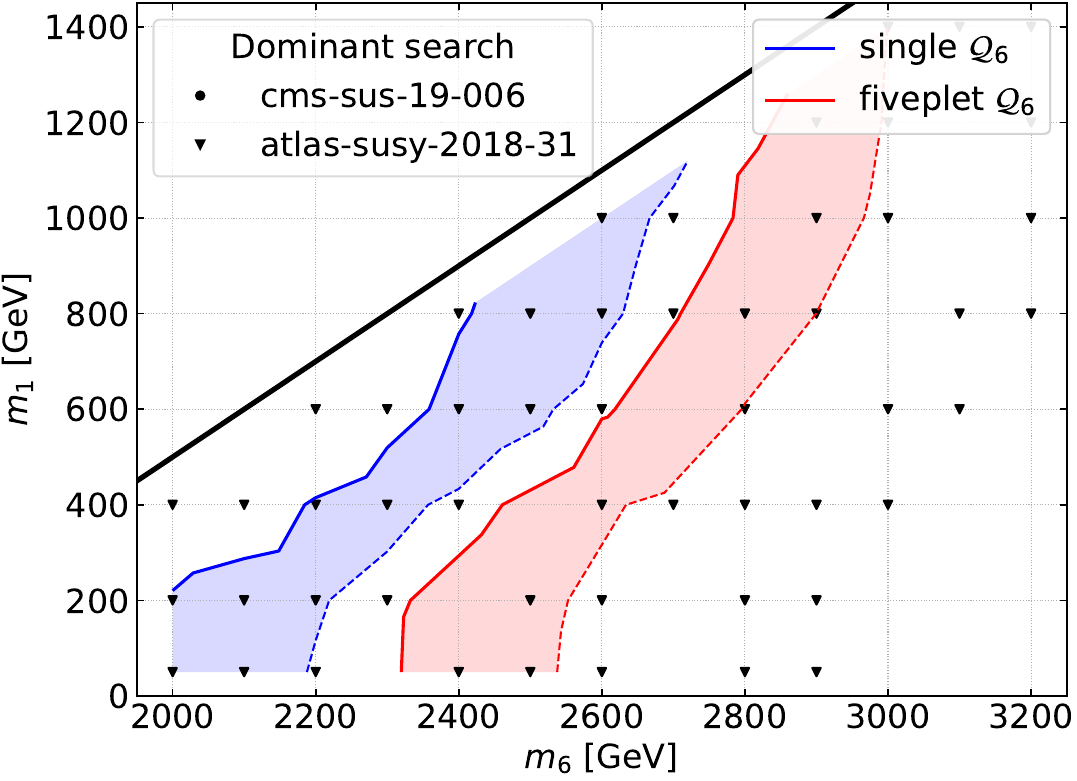}
    \includegraphics[width=0.45\linewidth]{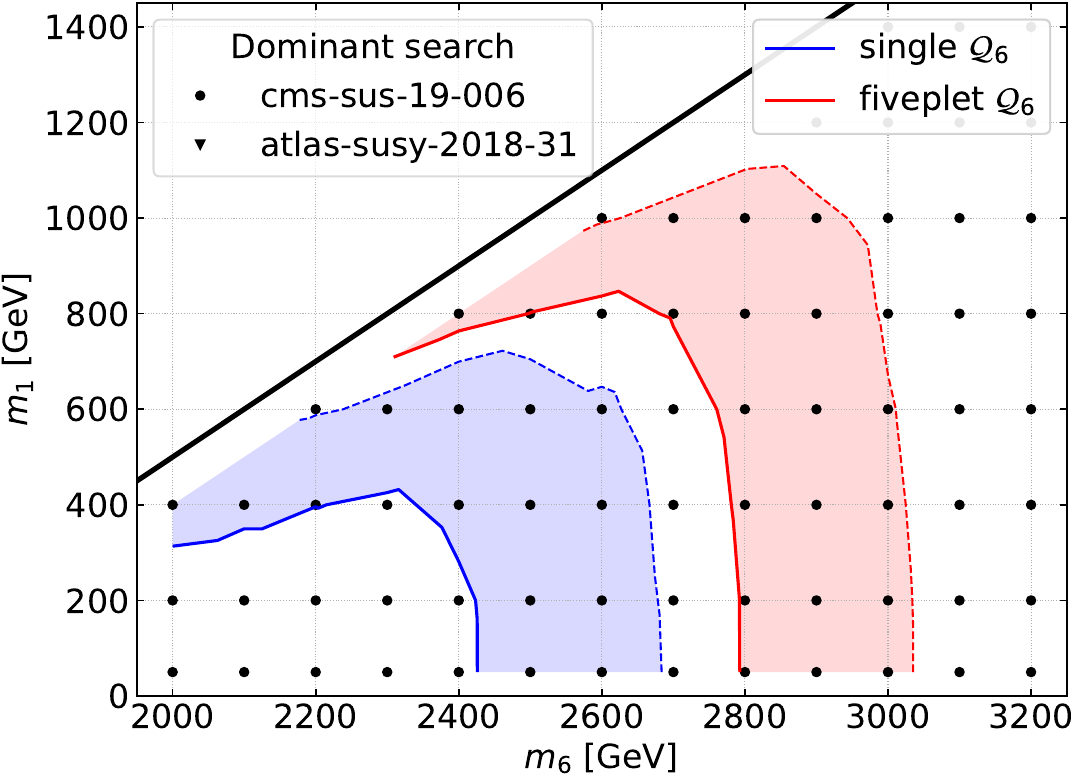}
    \caption{Same as \cref{fig:q6_DM_exclusion_details}, but after extrapolating the sensitivity to $3\,\mathrm{ab}^{-1}$ at fixed $\sqrt{s}=13$~TeV.}
    \label{fig:q6_DM_exclusion_HL_details}
\end{figure}

As an illustrative example, we focus on the bounds on the class-C1 missing-energy topology arising from the decay $\mathcal Q_6\to\pi_6\hat Q_1\to bb+E_T^{\rm miss}$, which we obtained by recasting the results of the \texttt{ATLAS-SUSY-2018-31}~\cite{ATLAS:2019gdh} and \texttt{CMS-SUS-19-006}~\cite{CMS:2019zmd} analyses in \cref{sec:LHC}. In this appendix we show the two contributions separately in order to clarify the origin of the shape of the combined contours when displayed in the $(m_6,m_1)$ plane in \cref{fig:q6_DM_exclusion}. The \texttt{CMS-SUS-19-006} analysis, which is designed for final states with large jet multiplicities, large $H_T$ and missing transverse momentum, provides broad coverage over much of the region where the decay $\mathcal Q_6\to\pi_6\hat Q_1$ is open. By contrast, the \texttt{ATLAS-SUSY-2018-31} analysis targets final states with several $b$-tagged jets and missing transverse momentum and becomes particularly relevant when the spectrum is more compressed and the detailed heavy-flavour structure of the final state plays a larger role. As expected, the current limits obtained from each search separately are shown in \cref{fig:q6_DM_exclusion_details}. The CMS bounds give the strongest reach over a large part of the low-$m_1$ and intermediate-$m_1$ region, especially for the full five-component sextet multiplet. The ATLAS limits, on the other hand, remain competitive near the diagonal threshold line and controls parts of the boundary of the combined exclusion. In \cref{fig:q6_DM_exclusion_HL_details} we show the corresponding extrapolations to $3~\mathrm{ab}^{-1}$ at fixed $\sqrt{s}=13$~TeV. Because the two analyses have different efficiencies across the $(m_6,m_1)$ plane, their luminosity extrapolations modify the shape of the envelope in a non-trivial way. In particular, the projected combined exclusion becomes more vertical over part of the plane, reflecting the fact that the limiting reach is then driven primarily by the sextet production rate rather than by the precise value of $m_1$ as long as the decay remains sufficiently far from the threshold.

\bibliographystyle{utphys}
\bibliography{literature}

\end{document}